\documentclass[11pt]{article}
\usepackage{articlestyle}
\RequirePackage{natbib}

\RequirePackage[colorlinks,pagebackref=false]{hyperref}
\hypersetup{
  colorlinks,
  linkcolor={red!50!black},
 linkcolor=magenta,
  citecolor={blue!50!black},
  urlcolor={blue!80!black}
  }

\newcommand{\revise}[1]{\textcolor{red}{#1}}
\newcommand{\han}[1]{{\textcolor{red}{Han: #1}}}
\newcommand{\drw}{^\mathrm{\scriptstyle drw}}

\newcommand{\source}{\mathcal{D}_S}
\newcommand{\target}{\mathcal{D}_T}
\def\fn{\footnote}

\numberwithin{equation}{section}

\begin{document}

\renewcommand{\baselinestretch}{1.1}\normalsize


\title{\Large \textbf{Transfer Learning for General Estimating Equations \\ with   Diverging  Number of Covariates }}
\author{
    Han Yan$^1$\,,~~~Song Xi Chen$^{1,2}$\\ 
    $^1$ Guanghua School of Management, Peking University, \\ 
    $^2$ School of Mathematical  Science, Peking University
}
\date{\today}
\maketitle

\vspace{-0.5cm}
\begin{abstract}

We consider statistical inference for parameters defined by general estimating equations under the covariate shift transfer learning. Different from the commonly used density ratio weighting approach, we undertake a set of formulations to make the statistical inference semiparametric efficient with simple inference.  It starts with re-constructing the estimation equations to make them Neyman orthogonal, which facilitates more robustness against errors in the estimation of two key nuisance functions, the density ratio and the conditional mean of the moment function. We present a divergence-based method to estimate the density ratio function, which is amenable to machine learning algorithms {including the deep learning}. To address the challenge that the conditional mean is parametric-dependent, we adopt a nonparametric multiple-imputation strategy that avoids regression at all possible parameter values. 
 With the estimated nuisance functions and the orthogonal estimation equation, the inference for the target parameter is formulated via the empirical likelihood without sample splittings.
We show that the proposed estimator attains the semiparametric efficiency bound, and the inference can be conducted with the Wilks' theorem. The proposed method is further evaluated by simulations and an empirical study on a transfer learning inference for ground-level ozone pollution.

\end{abstract}


\baselineskip=21pt

\renewcommand{\baselinestretch}{1.2}\normalsize


\section{Introduction}

The past decades have witnessed rapid development of statistical learning techniques in many fields of  applications.  
Most of the techniques 
rely on a commonly adopted model 
where the training and testing data are sampled from the same distribution. However, this homogeneity between the training and testing data is frequently violated in practice since 
diverse datasets are increasingly available. 
{An enduring challenge in statistical inference is to generalize an inference procedure from one data domain to another to achieve the goal of generalization and fully use of data information. 
} 

Transfer learning (TL) has become an active and promising area in dealing with distribution mismatch problems and has  achieved considerable success in a wide range of applications, such as computer vision (\citealp{Kulis2011}),  
bioinformatics (\citealp{hanson2020}) and precision medicine (\citealp{mo2021}). 
{See \cite{zhuang2020survey} for reviews.}

Suppose {that} the distributions for the source and target {samples} are $P_{\bX, Y}$
and $Q_{\bX, Y}$, respectively, where $\bX$ is the vector of covariates and $Y$ is the response or outcome. Two popular settings of TL that have been considered in the literature are 
posterior drift and covariate shift.  
The posterior drift TL assumes the marginal distributions of covariates are invariant, 
while the conditional distributions $P_{Y | \bX}$ and $Q_{Y | \bX}$ may differ. {On the other hand, covariate shift TL is for the situations where the marginal covariate distributions may differ,  
while a common conditional distribution is shared across the two domains. 
Recently, there has been a growing literature on statistical inference on the posterior drift TL, including {the} classification problems (\citealp{reeve2021adaptive}), {the} linear and generalized linear models (\citealp{li2022linear}), and {the} Gaussian graphical models (\citealp{li2023GGM}).  
Unlike the posterior drift,  the covariate shift TL has responses inaccessible to the target sample. Such a setting is well motivated by various real-world scenarios, where the same study is conducted with different covariate populations while the law that governs the input-output determination is kept across the domains.  
For instance, in medical data analysis (\citealp{guan2021}), covariate shift is reasonable for health records across different patients, and clinical {outcomes} are {usually} scarce because of ethical concerns. 
Despite its importance in applications, there {has been} a limited amount of literature on the statistical theory on the covariate shift TL relative to those for the posterior shift TL.

In this paper, we consider statistical inference on parameters defined  {via} general estimating equations (GEE) in the context of the covariate shift TL. The GEE is a general framework for semi-parametric inference, and is appealing for requiring less stringent distributional assumptions on the data, {and} yet can encompass a wide range of model structures and parameters. 
The goal is to efficiently estimate and make inference for a $p$-dimensional parameter $\btheta_0$ defined through $\E_Q \{ \bg(\bX, Y, \btheta_0) \} = \bzero$. 
Inference for $\btheta_0$ under this situation is more challenging than the conventional GEE problems because on one hand, $Y$ is available in the target sample. 
On the other hand,  directly using the sample from the source population $P$ leads to biased {estimates} since $P \neq Q$ in general. 

\subsection{Related works}
We review related works so as to situate our study within a broader context and discuss the gaps between existing results and the goal of this paper. 

\paragraph*{Covariate shift} The covariate shift, {as an important scenario of TL,} is also called the domain adaptation (\citealp{pan2010survey}) and has been investigated {in the machine learning literature}, such as \cite{gretton2009} and \cite{ryan2015semi}, with a focus to correct for {estimating bias} in the empirical risk minimization or model selection due to {the} covariate shift. The standard strategy adopted in the existing literature is the so-called importance reweighting with the density ratio between $Q_{\bX}$ and $P_{\bX}$; see \cite{kouw2019review} and the references therein. The covariate shift problem has also been {studied} from the perspectives of statistical methodologies and theories. \cite{Lei2021} studied conformal prediction under covariate shift. Non-parametric classification under covariate shift is explored in \cite{kpotufe2021} and non-parametric regression is investigated in \cite{ma2022}. 
\cite{cai2022-bandit} considered contextual multi-armed bandits under the covariate shift. In comparison, the semi-parametric inference under the covariate shift is less-explored.

\paragraph*{Missing data and causal inference}
The covariate shift TL 
is closely related to missing data problems and causal inference, 
since the assumption $P_{Y | \bX} = Q_{Y | \bX}$ is equivalent to the missing at random (MAR) condition. 
{The sample estimating equation employed in this work shares a similar form as the augmented inverse-probability weighted (AIPW) estimator (\citealp{robins_estimation_1994}) and its variants (e.g., \citealp{rotnitzky2012improved} and \citealp{chernozhukov2018}). Both the AIPW method and our proposed method require the estimation of the conditional mean function $\bbm(\bX,\btheta) = \E\{\bg(\bZ, \btheta) | \bX \}$.  A key distinction of the GEE considered in this paper from the aforementioned literature is that their estimand is commonly linear in $\bg(\bZ, \btheta)$, for example, the average treatment effect (ATE) problem 
corresponds to the estimating function $\bg(\bZ, \btheta) = Y - \theta$, while we are interested in more general cases where $\btheta$ may depend nonlinearly on $\bg(\bZ, \btheta)$, such as the quantile and quantile regression.
For the nonlinear cases, to estimate the conditional mean function $\bbm(\bX,\btheta)$, one has to regress $\bg(\bZ, \btheta)$ on $\bX$ repeatedly for each $\btheta$ during its optimization, which can be too computationally intensive to be practical, especially for some time consuming optimizations requiring thousands of iterations before convergence. 
}

{
Recently, \cite{chen2024}  considered the GEE problem with missing data and proposed a neural network based inverse probability weighting estimator. 
Compared with  \cite{chen2024}, our methods have two appealing advantages. One is being doubly robust in that our estimator is consistent if either the density ratio or the conditional mean function is consistently estimated. The other important feature is that we can conveniently employ Wilks' theorem for the inference of $\btheta_0$, while \cite{chen2024} has to resort to a Bootstrap method to facilitate the inference. More detailed comparisons are presented in Section \ref{sec: drw-theory}.}

\paragraph*{Empirical likelihood}
The empirical likelihood (EL) approach introduced in \cite{owen1988} has been demonstrated to be powerful for statistical inference of GEEs, for having appealing properties such as Wilks' theorem \citep{qin-lawless-94} and Bartlett correction \citep{chen-cui-07}. 
 When an unknown nuisance function is 
{present} in the estimation equations, \cite{hjort2009} and \cite{wang-chen-09} showed that asymptotically the empirical likelihood ratio statistic with a plugged-in estimate of the nuisance function can be weighted-sum-of-$\chi^2$ distributed, which is non-pivotal, and a bootstrap procedure has to be used to approximate the distribution {of the EL ratio}.  \cite{bravo2020} proposed a two-step procedure for empirical likelihood inference of semi-parametric models, employing a modified sample estimating function, 
which leads to an asymptotically $\chi^2$ distributed EL ratio statistic. {They also considered the GEE with missing data 
to illustrate their methodology. 
However, nuisance functions in \cite{bravo2020} are estimated with conventional kernel smoothing whose performance may deteriorate with increase of dimensionality. 
In our proposal, the nuisance functions are estimated in a more flexible way that accommodates modern ML algorithms. 
}

\subsection{Our contributions}
The investigation in this work contributes to several aspects. 

First, we construct a modified moment function for the GEE inference in the presence of covariate shift, which has the advantage of being Neyman orthogonal (\citealp{neyman1959}) {that permits elimination of the first-order effect of the nuisance function estimation,}
including a density ratio function $r(\bx)$ and a conditional moment function $\bbm(\bx, \btheta)$.

The second contribution is in proposing a novel estimation methods for the two nuisance functions, which both enable the use of flexible nonparametric tools, including the linear sieves and generic black-box machine learning algorithms.   
The density ratio function is estimated by a divergence minimization approach. 
The estimation of the conditional moment function $\bbm(\bx, \btheta)$ is more challenging since it requires estimating the nuisance for infinitely many $\btheta$. To tackle such a problem,
we employ a multiple imputation approach which bypasses the involvement of the parameter $\btheta$ and just needs to estimate the conditional density $p(y | \bX)$. Instead of the conventional kernel smoothing estimator, 
 novel estimation methods for the density ratio and the conditional density are presented, which can utilize a broad array of nonparametric methods.

Thirdly, by employing the EL method,  the proposed estimation is shown to be both doubly robust and semi-parametric efficient. Different from the double machine learning approach, the construction of the sample estimating function used for the EL does not require sample splitting.  Furthermore, the log EL ratio statistics admits the Wilks' theorem which greatly facilitates the inference. 
For comparison, we also investigate the theoretical properties of density ratio weighting (DRW) estimation, which is commonly adopted in covariate shift problems. It is found that the DRW not only requires more stringent conditions than the proposed method to attain the same asymptotic variance, but also make the EL ratio statistics asymptotically weighted $\chi^2$-distributed, which makes the subsequent inference tedious. 
We also discuss a growing dimension scenario, where the nuisance functions are estimated with the deep neural networks to circumvent the curse of dimensionality. 

\subsection{Orginization}
The paper is organized as follows. Section \ref{sec: setup} describes the setup of the GEE problem with the covariate shift TL. In Section \ref{sec: m1} the orthogonal moment equations are constructed. Section \ref{sec: nuisance} discusses the estimation of the two nuisance functions and establishes their theoretical properties. The inference with the EL is presented in Section \ref{sec: EL}, where the scenario with a growing dimension is also investigated. In \label{sec: relate} we discuss comparisons between our study and some related works. Section \ref{sec: sim} and \ref{sec: case} report numerical experiment results and a case study on ground-level ozone pollution, respectively. Finally, concluding discussions are given in Section \ref{sec: discuss}.

\section{Notation and problem setup}\label{sec: setup}
We first introduce notations used {in this study}. 
We use $\mathds{1}(\mathcal{A})$ as the indicator function of an event $\mathcal{A}$. 
For any vector $\bv = (v_1, \cdots, v_d)\t$, let $\bv^{\otimes 2} = \bv \bv\t$ and $\norm{\bv}_p$ denote its $L^p$ norm. 
For a function $f: \mathcal{X} \to \mathbb{R}$, its supreme is denoted by $\norm{f}_{\infty} = \sup_{\bx \in \mathcal{X}}f(\bx)$, and its $L_p$-norm under a distribution $F$ is denoted by $\norm{f}_{L_p(F)} = (\E_F |f(X)|^p)^{1/p}$ for any $p \geq 1$. 
For two sequences of positive numbers $ \{a_{n}\} $ and $ \{ b_{n}\} $, we write $a_{n} \lesssim  b_{n}$ if there exists a positive  constant $C$ such that  $a_{n} \leq C b_{n}$. 

Suppose a source sample $\source$ has $n$ independently and identically distributed (i.i.d.) observations $\bZ_{1},\dots, \bZ_{n}$ from a source population  $\bZ \sim P= P_{\bX} \times P_{Y|\bX}$ where $\bZ_{i} = (\bX_{i}\t, Y_{i})\t$ consists of  a $d$-dimensional covariate $\bX_{i}$ and  a response/label $Y_{i}$. In this study, we take the response variable as a scalar, 
 as the case of multivariate responses can be readily extended.  The target population is $\bZ \sim Q = Q_{\bX} \times Q_{Y|\bX}$. Observations of the target sample $\target$ are $\bX_{n+1},\dots, \bX_{N}$ with $N = n+m$, while the responses $Y_{i}$  in $\target$ are \textit{not} accessible. We introduce a binary variable $\delta$ to indicate whether the data is drawn from the source ($\delta = 0$) or the target ($\delta = 1$) population.
Let $\tau= \P(\delta = 1)$ 
denote the proportion of target observations in {the entire} $N$ observations, which is approximated by $m /N$.     

Let $\btheta = (\theta_{1},\dots, \theta_{p})\t$ be a $p$-dimensional parameter taking values in $\bTheta \subset \mathbb{R}^p$. For a set of   estimating equations $\{g_{i}(\bZ, \btheta)\}_{i=1}^r$, the true parameter $\btheta_{0} \in \bTheta$ of the target population is {identified} by the moment condition
\begin{equation} \label{eq: EE}
     \E_{Q}  \{ \bg(\bZ,\btheta_0)\} =\bzero,
\end{equation} 
where $\bg(\bZ, \btheta)=\left(g_{1}(\bZ, \btheta), \ldots, g_{r}(\bZ, \btheta)\right)\t$  with $r \geq p$, which is necessary for identifying $\btheta_{0}$.
If $\btheta_0$ depends on $Q_Y$, then without further distributional conditions it is impossible to identify $\btheta_0$ using the observed data due to the missingness of $Y$ in $\target$.  

Following the standard setting in the covariate shift literature, 
we assume that the conditional distributions $P_{Y|\bX} = Q_{Y|\bX}$ so that the information of $Y$ can be transferred from the source sample $\source$ to the target sample $\target$, while the covariate distributions $P_{\bX}$ and $Q_{\bX}$ can differ.  
Our interest is the inference on $\btheta_0$ with the {combined} sample $\mathcal{D} = \source \cup \target$ under the covariate shift. 
The following conditions are required for the sample and target populations. 

\begin{con}\label{con: dist}
(i) The covariate distributions $P_{\bX}$ and $Q_{\bX}$ are absolutely continuous with densities $p_0(\bx)$ and $q_0(\bx)$ supported on $\mathcal{X}$, where $\mathcal{X} \subset \mathbb{R}^d$ is compact. 
(ii) The conditional distributions 
$P_{Y | \bX = \bx} = Q_{Y | \bX = \bx}$ for every $\bx \in \mathcal{X}$. 
\end{con}

\begin{con}\label{con: moment function}

(i) The parameter $\btheta_0 \in \text{int}(\Theta)$ is the unique solution to the moment condition $\E_Q\{\bg(\bZ, \btheta) \} = 0$. (ii) $\E_Q\{\sup_{\btheta \in \Theta}\norm{\bg(\bZ, \btheta)}_2^{\alpha} \} < \infty$ for some $\alpha > 2$. (iii) The eigenvalues of $\E_Q\{ \bg(\bZ, \btheta)^{\otimes 2} \}$ are bounded away from zero and infinity. (iv) $\bg(\bz, \btheta)$ is continuously differentiable in a neighborhood $\mathcal{N}$ of $\btheta_0$ with $\E_Q\{\sup_{\btheta \in \cN} \norm{\partial \bg(\bZ, \btheta) / \partial \btheta\t}_2\} < \infty$, and $\E_Q\{\partial \bg(\bZ, \btheta_0) / \partial\btheta\}$ is of full rank.

\end{con}

Condition \ref{con: dist} summarizes assumptions about the sample and target populations under the setting of covariate shift,
where 
Condition \ref{con: dist} (ii)  is necessary for inferring information regarding the target domain from the source domain with accessible responses. Such a condition is also required in semi-supervised learning problems (e.g., \citealp{ryan2015semi}). However, more challenging than the semi-supervised learning setup, we do not assume that $P_{\bX}$ and $Q_{\bX}$ are the same. 
Condition \ref{con: moment function} for the estimating functions are standard regularity conditions in the literature of general estimation equations (e.g., \citealp{newey-smith-04}).

\section{Orthogonal moment functions}
\label{sec: m1}

To address the problems caused by the covariate shift, the most common existing method 
is via a density ratio weighting (DRW) approach, see \cite{sugiyama2007} for an empirical risk minimization, \cite{Lei2021} for conformal predictions, and \cite{ma2022} for the kernel ridge regression under the covariate shift.  
However, 
 as will be revealed shortly,  the DRW method may not be suitable for the inference of the GEEs under the covariate shift. We will propose to modify the DRW moment functions into an orthogonal moment function that is more robust against the estimation error of the density ratio function.  

Let 
$r_{0}(\bx) = q_{0}(\bx) / p_{0}(\bx)$ be the density ratio of $Q_{\bX}$ and $P_{\bX}$. 
Using $r_{0}(\bX)$ to weigh for $\bZ$ from the source population, it holds that 
\begin{equation*} 
    \E_{P}  \{r_{0}(\bX) \bg(\bZ, \btheta) \} = \E_{Q} \left\{\bg(\bZ, \btheta) \right\},   
\end{equation*}
for any $\btheta \in \bTheta$.
The above relations reveal the central role of the density ratio function in the identification of $\btheta_0$ using the fully observed source sample. 
With a consistent $\widehat{r}(\bx)$, we can obtain an estimate $\wh{\btheta}\drw$ from the following density ratio weighting (DRW) moment function  
\begin{equation}\label{eq: drw-ef}
    \tilde{\bg}(\bZ_{i}, \btheta,\widehat{r}) =
    \widehat{r}(\bX_{i}) \bg(\bZ_{i}, \btheta)  ~~~\text{for} ~i = 1,\dots, n,
\end{equation}
with either the empirical likelihood or the generalized method of moments approach. 

While being the most popular and natural strategy for tackling the covariate shift, the DRW method for the GEE problem has 
several 
 drawbacks. First, the {accuracy} 
 of $\wh{\btheta}\drw$ crucially depends on that of $\wh{r}$, which may {not be high quality} when $r_0$ has a complex structure or the model of $r_0$ is misspecified. A more important problem arises in the 
 inference 
  as the estimation error of $\wh{r}$ may have a 
 first-order effect on  $\wh{\btheta}\drw$.  
 This is because the asymptotic distribution of $\wh{\btheta}\drw$ depends 
 on 
 that of the partial sum $n^{-\frac{1}{2}}\sum_{i=1}^n   \widetilde{\bg}(\bZ_{i},\btheta_{0},\widehat{r})$, which can be decomposed as
\[ 
\frac{1}{\sqrt{n}} \sum_{i=1}^n  \widetilde{\bg}(\bZ_{i},\btheta_{0},\widehat{r}) =   \frac{1}{\sqrt{n}} \sum_{i=1}^n   \widetilde{\bg}(\bZ_{i},\btheta_0,r_{0}) + \frac{1}{\sqrt{n}} \sum_{i=1}^n \bg(\bZ_i, \btheta_0)  \{ \widehat{r}(\bX_i) - r_0(\bX_i)\} := T_n + R_n,
\]
where $T_n$ is usually 
 asymptotically Gaussian, 
However, the second term $R_n$, which gathers effects of the plugged-in estimator $\wh{r}$, may not have a weak limit, 
  especially when $\wh{r}$ is obtained from some black-box machine learning methods. 
As shown 
 in Section \ref{sec: drw-theory}, 
 $\wh{\btheta}\drw$ requires quite strong conditions to be asymptotically normal.  
 Even under such a case,  Theorem \ref{thm: drw} shows that the EL ratio statistic using \eqref{eq: drw-ef} as moment functions has a weighed-$\chi^2$ limiting distribution, whose quantiles require a Bootstrap procedure to approximate. 
 See also \cite{hjort2009}  on the weighted-$\chi^2$ phenomenon associated with the EL 
  with plugged-in nuisance function estimators.

To alleviate the effect of estimation {error} in the nuisance function $r_0(\bx)$,  
we opt for adjusting the sample moment function $\tilde{\bg}(\bZ_i, \btheta, \wh{r})$ by employing the first-order correction 
advocated in the semiparametric literature such as \cite{newey1994asymptotic}. 
To illustrate the idea, 
let $\bm{\mu}( \btheta,\wh{r}) = \E_P\{\bg(\bZ, \btheta, \wh{r}) \}$ and consider  the first-order von Mises expansion 
\begin{align} 
\bm{\mu}(\btheta, r_0) = \bm{\mu}(\btheta,\wh{r}) + \int \bm{\psi}(\bz,\btheta,\wh{r}) d P(\bz) + R_2(\wh{r}, r_0),  
\label{eq: von-mises}
\end{align}
where $\bm{\psi}(\bz, \btheta, \wh{r}) $ is the pathwise derivative of 
$\bm{\mu}(\btheta,r)$ at $\wh{r}$ and $R_2(\wh{r}, r_0)$ is the 
reminder term. The 
 expansion suggests that $\bm{\psi}(\bz,\btheta, \wh{r})$ represents the plugged-in effect of $\wh{r}$, 
 and the bias of the weighted moment function $\tilde{\bg}(\bZ_i, \btheta, \wh{r})$, namely $\bm{\mu}(\btheta,\wh{r}) - \bm{\mu}(\btheta, r_0)$, can be corrected by adding back 
$\int \bm{\psi}(\bz, \btheta, \wh{r}) d P(\bz)$. As will be shown 
 in Theorem \ref{thm1},  the pathwise derivative satisfies 
\begin{align}
\int \bm{\psi}(\bz,\btheta,\wh{r}) d P(\bz) = \E_Q\{\bbm_0(\bX, \btheta) \} - \E_P\{\bbm_0(\bX, \btheta) \wh{r}(\bX)\}. \nn 
\end{align}
With an estimated conditional mean function $\wh{\bbm}(\bx, \btheta)$ {using the source sample $\source$},
the above quantity can be approximated by $\sum_{i=1}^N  \tilde{\bm{\psi}}(\bW_i, \btheta, \wh{\bfeta})$, where $\bW_i = (\bZ_i, \delta_i)$, 
 $\wh{\bfeta} = (\wh{r}, \wh{\bbm})$ and 
\[
\tilde{\bm{\psi}}(\bW_i, \btheta, \wh{\bfeta}) = \frac{\delta_i}{\tau} \wh{\bbm}(\bX_i, \btheta) - \frac{1-\delta_i}{1-\tau} \wh{\bbm}(\bX_i, \btheta)\wh{r}(\bX_i) 
\]
for $i = 1, \cdots, N$. 
Adding the adjustment 
$\tilde{\bm{\psi}}(\bW_i, \btheta, \wh{\bfeta})$ to the weighted moment function $\tilde{\bg}(\bZ_i, \btheta, \wh{r})$ leads to the following 
 moment function 
\begin{equation}\label{eq: orthogonal moment}
     \bPsi(\bW_i, \btheta, \wh{\bfeta}) =   \frac{1-\delta_i}{1-\tau} \wh{r}(\bX_i) \{  \bg(\bZ_i,\btheta) -\wh{\bbm}(\bX_i, \btheta) \} +   \frac{\delta_i}{\tau} \wh{\bbm}(\bX_i, \btheta).
\end{equation}

A direct calculation verifies that $\E\{  \bPsi(\bW, \btheta_0, \bfeta_0) \} = \bzero$ at $\bfeta_0 = (r_0, \bbm_0)$, 
which implies that $ \bPsi(\bW_i, \btheta, \wh{\bfeta})$ is 
a valid moment function for identifying $\btheta_0$. 
We now establish the key aspects 
regarding the proposed moment function. 
Let $F$ be the mixture of the source and target distributions. 
{To derive the pathwise derivative 
of the functional $r \mapsto \E_F\{\widetilde{\bg}(\bW,\btheta, r) \}$, 
let  $ \{ F_{\tau}, \tau \in [0,1)\} $  be a collection of regular parametric 
} 
submodels satisfying $F_{0} = F$ and the mean-square differentiability condition (see, e.g., \citealp{van2000asymptotic}). 
The true nuisance function under the submodel $F_{\tau}$ is denoted as $\bfeta(F_{\tau})$ such  that $r(F_{\tau})$ is the true covariate density ratio under $F_{\tau}$

\begin{thm}
    \label{thm1}
    Under Conditions \ref{con: dist} and \ref{con: moment function},  
    the following results hold. 
\noindent {(i)}
For any $\btheta \in \Theta$, 
 \begin{equation}
    \frac{\partial}{\partial \tau} \E_{F} \{ \widetilde{\bg}(\bW,\btheta, r(F_{\tau}))\}\Big\vert_{\tau = 0}=   \E_{F} \left\{ \bvphi(\bW, \btheta, \bfeta_{0})  S_{0}(\bW) \right\}, 
\end{equation}   
where $\bfeta_{0}(\bx, \btheta) = (r_{0}(\bx), \bbm_{0}(\bx,\btheta))$ 
and 
    $\bvphi(\bw, \btheta, \bfeta) = \frac{\delta}{p} \bbm(\bx, \btheta) - \frac{1-\delta}{1-p}{r}(\bx) \bbm(\bx, \btheta)$.

\noindent  {(ii)}
Let $\bPsi(\bw, \btheta, \bfeta) =  \widetilde{\bg}(\bw, \btheta, r) +  \bvphi(\bw, \btheta, \bfeta)$ or equivalently,  
\begin{align}
        \bPsi(\bw, \btheta, \bfeta) =  \frac{1-\delta}{1-p} r(\bx) \{  \bg(\bz,\btheta) - \bbm(\bx, \btheta) \} +   \frac{\delta}{p} \bbm(\bx, \btheta),
       \label{eq: Psi moment}
   \end{align} 
then 
      $\frac{\partial}{\partial \tau} \E_{F} \{ \bPsi(\bW,\btheta_0, \bfeta(F_{\tau}))\}\Big\vert_{\tau = 0} = \bzero$.  

\noindent {(iii)}  For any candidate $\bfeta(\bx, \btheta)  = (r(\bx),\bbm(\bx,\btheta))$,  
 $$
 \norm{\E_{F} \{ \bPsi(\bW,\btheta_{0}, \bfeta)  \}}_1 \leq  \norm{r(\bX) - r_0(\bX)}_{L_2(P_{\bX})} (\sum_{j=1}^r \norm{m_j(\bX, \btheta) - m_{0j}(\bX, \btheta)}_{L_2(P_{\bX})}). 
 $$
\end{thm}
Theorem \ref{thm1} (i) shows the pathwise derivative function of $\E_{F} \{ \widetilde{\bg}(\bW,\btheta, r(F_{\tau}))\}$ is $  \bvphi(\bw, \btheta, \bfeta)$, reflecting the local effect the density ratio $r(F_\tau)$ on $\E_{F} \{ \widetilde{\bg}(\bW,\btheta, r(F_{\tau}))\}$. The property $\frac{\partial}{\partial \tau} \E_{F} \{ \bPsi(\bW,\btheta_0, \bfeta(F_{\tau}))\}\Big\vert_{\tau = 0} = \bzero$ in Theorem \ref{thm1} (ii) is the so-called Neyman orthogonality 
(\citealp{neyman1959}, \citealp{chernozhukov2018}), which means 
the proposed sample moment function $\bPsi(\bW, \btheta, \bfeta)$ is orthogonal to the nuisance functions. 
Based on such a property, perturbing the nuisance function $\bfeta$ locally around 
$\bfeta_0$ does not have the first-order effect on $\E \{ \bPsi(\bW,\btheta_{0}, \bfeta)  \}$. 
The Neyman orthogonality is an important notion in semi-parametric inference as {it enables the estimating function to be locally insensitive to the nuisance function.} Compared with the debiased machine learning (DML) proposed by \cite{chernozhukov2018} which also utilized Neyman orthogonal moments, the problem considered here is more challenging, as $\bbm(\bx, \btheta)$ is parameter-dependent. 
Theorem \ref{thm1} (iii) reveals that the bias of the moment functions $\E \{ \bPsi(\bW,\btheta_{0}, \bfeta)  \}$ is 
bounded by the \textit{product} of the $L_2$ norms of the estimation errors of the two nuisance functions, which is related to the double robustness property introduced by \cite{robins_estimation_1994} for the augmented inverse probability weighting (AIPW) estimator. 

\section{Estimation of nuisance functions}
\label{sec: nuisance}

Given the important roles played by the two nuisance functions $r(\bx)$ and $\bbm(\bx, \btheta)$, 
this section proposes estimators of the two nuisance functions and discuss their theoretical properties.  

\subsection{Density ratio estimation}\label{sec: m2}

{We first present estimators to the density ratio $r$. Conventional approaches, 
such as the kernel smoothing or the classification-based methods, typically estimate the density functions of the target domain (numerator) and the source domain (denominator), respectively, to form the ratio estimator. However, such density ratio estimators  
can be quite unstable when the dimension is large or the denominator density is close to zero.  We take an 
approach that \textit{directly} estimate the density ratio based on the dual characteristic of the  $\phi$-divergence, which can be 
solved via an empirical risk minimization problem and can 
accommodates a variety of machine learning algorithms. 
}

For  {any two distributions $P$ and $Q$ with densities $p_0$ and $q_0$ and suppose that $P$ is absolutely continuous with respect to $Q$,} 
their $\phi$-divergence is 
\begin{equation}\label{eq: D phi}
    D_{\phi}(Q \| P) = \int_\mathcal{X} \phi \left( \frac{q_{0}(\bx)}{p_{0}(\bx)} \right)  p_{0}(\bx)  d \bx,
\end{equation}
where  $\phi: \mathbb{R}_{+} \to \mathbb{R}$ is a convex and lower semicontinuous function. Different choices of $\phi$ lead to  different divergences, 
such as the KL divergence for $\phi(u) = u\log u$, the squared Hellinger distance for $\phi(u) = (\sqrt{u} - 1)^2$, and the Pearson's $\chi^2$-divergence for $\phi(u) = (u-1)^2$. See \cite{sason2016} for more examples. {Moreover, the class of the Cressie-Read power divergence can be represented as $\phi$-divergences as shown in \cite{maji2019}. } 

Let $\phi_{*}(v) = \sup_{u \in \mathbb{R} } \{ uv - \phi(u) \}$ be the Fenchel dual function of $\phi$. {The dual representation theorem (\citealp{rockafellar1997})  implies that
\begin{align}
      D_{\phi}(Q \| P) & = \sup_{f: \cX \to \text{Dom}(\phi_*) }\left\{ \E_Q (f)  - \E_P\left( \phi_{*}(f)\right) \right\} \nn  \\ 
      & =  \E_Q (f_0)  - \E_P\left( \phi_{*}(f_0)\right),
       \label{eq: dual of phi} 
\end{align}
where $\text{Dom}(\phi_*)$ denotes the domain of $\phi_*$, and the supreme is attained at $f_{0}(\bx) = \phi'\left( \frac{q_0(\bx)}{p_0(\bx)}\right) = \phi'\left( r_0(\bx) \right)$.}
For each $\phi$-function, we define 
\begin{equation}
    \ell_{1, \phi}(r) = \phi_*\{\phi'(r)\}~~\text{and}~~ \ell_{2, \phi}(r) = \phi'(r).
\end{equation}
Then the relationship \eqref{eq: dual of phi} induces an identification condition for the density ratio $r_0$ as presented in the following lemma. 
\begin{lem}
    For any convex and lower semicontinuous function $\phi: \mathbb{R}_{+} \to \mathbb{R}$, 
    the true density ratio  satisfies 
\begin{align}
    r_0 = \argmin_{r \in {\mathcal{F}}} L_{\phi}(r) ~~\text{with}~~ L_{\phi}(r) = \E_P\{\ell_{1, \phi}(r) \} - \E_Q\{\ell_{2, \phi}(r) \},
    \label{eq: drf cri}
\end{align}    
where {the candidate class $\cF$ is any class of nonnegative functions that contains $r_0$.} 
\label{lem: dual of phi}
\end{lem}  
The proof of the above lemma is presented in the supplementary material (SM). {For each given} 
$\phi$ function, $r_0$ can be uniquely determined from the population objective function \eqref{eq: drf cri}.  Table \ref{tab: ex of phi} lists some examples of commonly used divergence, along with the corresponding Fenchel conjugate function $\phi_*$ and the objective functions $\ell_{1, \phi}$ and  $\ell_{2, \phi}$.

\begin{table}[ht]
\caption{Examples of $\phi$-divergence, the associated Fenchel conjugate and the objective 
functions.}
    \centering
    \begin{tabular}{c c c c c }
    \hline 
    Divergence & $\phi(u)$ & $\phi_*(v)$ & $\ell_{1, \phi}(r)$ & $\ell_{2, \phi}(r)$  \\
    \hline
     Kullback-Leibler &  $u \log(u)$ & $\exp(v-1)$ & $r$ & $\log(r) + 1$ \\
     Reverse KL & $ - \log(u)$ & $-1 - \log(-v)$ & $\log(r) + 1$ & $-r^{-1}$ \\ 
     Pearson $\chi^2$ & $(u-1)^2$ & $v^2 / 4 + v$ & $r^2 -1$ & $2(r-1)$\\ 
     Squared Hellinger & $(\sqrt{u} -1 )^2$ & $v/(v-1)$ & $r^{\frac{1}{2}} - 1$ & $1 - r^{-\frac{1}{2}}$ \\
     \hline
    \end{tabular}
    \label{tab: ex of phi}
\end{table}

With the two samples from $P$ and $Q$, the density ratio $r_0$ can be estimated with the sample objective function obtained by replacing the expectations in \eqref{eq: drf cri} with the corresponding empirical averages. 
The function class $\cF$
in  \eqref{eq: drf cri} is required to contain the true density ratio $r_0$, whose functional form
is generally unknown in practice. 
As a practical surrogate of $\cF$, we use a candidate class $\cF_N$ which may not exactly contain $r_0$ but has the universal approximation ability as described in Condition \ref{con: app r}. Such a requirement can be satisfied by the linear sieves and numerous machine learning methods. 
The density ratio estimator $\widehat{r}$ %
is given by  
\begin{equation}
    \widehat{r}=  \argmin_{r \in \mathcal{F}_N} \left\{ 
    \frac{1}{n} \sum_{i=1}^n  \ell_{1, \phi}\{r(\bX_i)\} - \frac{1}{m}\sum_{i= n+1}^N\ell_{2, \phi}\{r(\bX_i)\} 
    \right\}. 
     \label{eq: dre}
\end{equation}
It is noted that 
we not only obtain the  estimator 
$\wh{r}$, {but also an 
estimate of the divergence $D_{\phi} (Q\| P)$ by the sample objective function with $\wh{r}$, which 
reveals the discrepancy of the source and the target populations.}  
The procedure applies to any $\phi$-divergence 
introduced by different choices of $\phi$.
For example,  
choosing $\phi(u) = u\log(u)$ corresponds to the  KL-divergence in Table \ref{tab: ex of phi} such that  
\begin{equation}\label{eq: r-kl}
     \widehat{r}  = \argmin_{r \in \mathcal{F}_N} \left\{   \frac{1}{n} \sum_{i=1}^n r(\bX_{i})  - \frac{1}{m}\sum_{i= n+1}^N \log\{r(\bX_{i})\} \right\}.
\end{equation} 

{Since we have formulated the estimation for $r_0$ into an empirical risk minimization problem, a variety of computational methods for the optimization can be applied. If the candidate function space $\cF_N$ is convex, then problem \eqref{eq: r-kl} is a convex programming problem, as was demonstrated in \cite{nguyen2010} for $\cF_N$ being the reproducing kernel Hilbert space (RKHS). For more general nonparametric function classes such as the deep neural networks (DNN), the optimization can be conducted via efficient computational algorithms such as the stochastic gradient descent.  We advocate the use of the DNNs in the scenario of large-scale data, since the DNNs are more amenable to parallel computations (\citealp{goodfellow2016deep}). Furthermore, compared with the RKHS employed in \cite{nguyen2010}, the DNNs also enjoyed the advantage of adaptivity to unknown low-dimensional structures of the underlying function, thus mitigating the curse of dimensionality, as will be discussed in Section \ref{sec: hd}.
}

Now we study the $L_2$-estimation error of the proposed density ratio estimator by first presenting the results for the H\"{o}lder function class.  

\begin{definition}
Let $\beta = \lfloor \beta \rfloor + r > 0, r \in (0,1]$ where  
$\lfloor \beta \rfloor$ denotes the largest 
integer strictly smaller than $\beta$. 
    For a finite constant $B > 0$ and a compact region $\mathcal{X} \subset \mathbb{R}^d$, the H\"{o}lder function class  
    \[
    \mathcal{H}^\beta(\mathcal{X}, B) = \left\{ f: \mathcal{X} \to \mathbb{R}, \max_{\norm{\balpha}_1\leq \lfloor \beta \rfloor} \norm{\partial^{\balpha}f}_{\infty} \leq B, ~\max_{\norm{\balpha}_1 =  \lfloor \beta \rfloor}\sup_{\bx_1 \neq \bx_2 \in \mathcal{X}} \frac{\partial^{\balpha}f(\bx_1) - \partial^{\balpha}f(\bx_2)}{\norm{\bx_1 - \bx_2}_2^r}  \leq B\right\},
    \]
    where $\partial^{\balpha} = \partial^{\alpha_1}\cdots\partial^{\alpha_d}$ with $\balpha = (\alpha_1, \cdots, \alpha_d)\t \in \mathbb{N}^d$ and $\norm{\balpha}_1 = \sum_{i=1}^d \alpha_i$.
\end{definition}

\begin{con}\label{con: r0}
There exist constants $B_1 > 0$ and $\beta_1 \geq 1$ such that the target function $r_0 \in \mathcal{H}^{\beta_1}(\mathcal{X}, B_1)$. 
\end{con}

\begin{con}\label{con: app r}
 Let the pseudo-dimension (see \citealp{pollard1990}) of $\mathcal{F}_N$ be $\pdim(\mathcal{F}_N)$,  then 
 (i)  $\pdim(\mathcal{F}_N) \log(N) = o(N)$; and (ii) there exists a constant $c_2 > 0$ such that for large enough $n$,
 $\inf_{r \in \mathcal{F}_N}\norm{r - r_0}_{\infty} \leq c_2 \pdim(\mathcal{F}_N)^{-\frac{\beta_1}{d}}.$ (iii) There exists a positive constant $M_1$ such that $\norm{r}_\infty \leq M_1$ and $\norm{\ell_{i, \phi}''(r)}_{\infty} \leq M_1 $ for $i = 1, 2$ and for every $r \in \cF_N$. 
\end{con}

The above two conditions are imposed on the true density ratio $r_0$ and the candidate function class $\cF_N$, respectively. Condition 
\ref{con: r0} characterizes the smoothness of $r_0$, as commonly imposed in nonparametric function estimation. Condition \ref{con: app r} restricts the complexity of  $\cF_N$ and assumes the approximation error $\inf_{r \in \mathcal{F}_N}\norm{r - r_0}_{\infty}$ converges to $0$ with the increase of the pseudo-dimension $\pdim(\cF_N)$. Such a condition can be satisfied by various nonparametric function classes, including the linear sieves (\citealp{chen2007sieve}), such as the splines and the wavelets, and also the many machine learning methods, for example, the deep neural networks (\citealp{jiao2022}).
{Condition \ref{con: app r} (iii) ensures that every function $r$ in $\cF_N$ as well as the second derivative of $\ell_{i, \phi}(r)$ are bounded by $M_1$,  which can be practically achieved by a truncation operation.} 
With the above conditions, we have the following result for the estimation error of the proposed estimator $\hat{r}$. 
{To quantity the  estimation performance, we define empirical $L_2$ error of $\wh{r}$ as
\begin{equation}\label{eq: error of r}
\cE_N(\wh{r}) = [N^{-1}\sumiN \left\{ \wh{r}(\bX_i) - r_0(\bX_i)\right\}^2]^{1/2}. 
\end{equation}
}
{
\begin{thm}\label{thm3}
Under Conditions \ref{con: dist}, \ref{con: r0}, and \ref{con: app r}, there exists a positive constant $C_1$ such that  with probability at least $1 -2 e^{-t}$, for $N$ large enough and any $t > 0$, 
\begin{align}
  &  \cE_N(\wh{r}) \leq  C_1 \left(\sqrt{\frac{\pdim(\mathcal{F}_N) \log(N)}{N}} +
   \inf_{r \in \cF_N} \norm{r - r_0}_{\infty}+ \sqrt{\frac{t}{N}} \right).
    \label{eq: non-asy bound}
\end{align}
\end{thm}
}
{The theorem provides the non-asymptotic estimation error bound for $\wh{r}$. The proof of the theorem is presented in Section B.2 of the SM, which is built on a scale-sensitive localization theory (\citealp{koltchinskii2011}) to derive the tight bounds on the estimation errors. In the proof of Theorem \ref{thm3}, we also show that the population  $L_2$ error of $\wh{r}$ can be bounded by half of the right-hand side of \eqref{eq: non-asy bound} with high probability,  as a consequence of Talagrand's concentration 
The first two terms of the bounds in \eqref{eq: non-asy bound} correspond to the stochastic error determined by the complexity $\pdim(\cF_N)$ and the approximation error $\inf_{r \in \cF_N} \norm{r - r_0}_{\infty}$, respectively. Under Condition \ref{con: app r} (ii), the second term can be bounded by $\pdim(\cF_N)^{-\frac{\beta_1}{d}}$. Therefore, there is a trade-off between the first two terms on the right-hand side of \eqref{eq: non-asy bound} with respect to the increase of the complexity of the candidate class $\cF_N$.  
Balancing the first two terms, it can be seen that $\pdim(\mathcal{F}_N) = O(N^{-\frac{d}{2\beta_1 + d}})$ is the optimal choice of pseudo-dimension up to some $\log(N)$ factor.}
In practice, while the underlying smoothness $\beta_1$ is generally unknown, we can specify the optimal $\pdim(\mathcal{F}_N)$ with the cross-validation method. With such the optimal specification of $\pdim(\mathcal{F}_N)$, the following convergence rate of $\wh{r}$ can be obtained.

\begin{coro}\label{thm: conv-r}
Under Conditions \ref{con: dist}, \ref{con: r0}, and \ref{con: app r}, and taking 
 $\pdim(\mathcal{F}_N) = O(N^{-\frac{d}{2\beta_1 + d}})$, 
we have 
\[
  \cE_N(\wh{r}) = O_p\left( N^{-\frac{\beta_1}{2\beta_1 + d}} \log^{\frac{1}{2}}(N)\right). 
\]
\end{coro}

Theorem \ref{thm: conv-r} establishes the convergence rate of the proposed density ratio estimator. We note that the  $N^{-\frac{\beta_1}{2\beta_1 + d}}$ 
is the minimax lower bound for the density estimation problem as shown in \cite{stone1982optimal} and \cite{yang1999minimax}. However, the density ratio estimation is a harder problem than the density estimation as the former is a two-sample problem. In the next theorem, the minimax lower bound for the density ratio estimation is derived.
\begin{thm}\label{thm: minimax for r}
Let $\mathcal{M}^d(\beta_1, B_1) = \left\{ (\mathbb{P}, \mathbb{Q}): d \Q / d\P = r_0 \in \mathcal{H}^{\beta_1}(\mathcal{X}, B_1) \right\}$, then there exists a positive constant $c_1$  such that 
\[
\inf_{\tilde{r} } \sup_{\substack{(\mathbb{P}, \mathbb{Q} ) \in \mathcal{M}^d(\beta, B) }}  \E\{\norm{r - r_0}_{L_2(P)} \}  \geq c_1 N^{-\frac{\beta_1}{2\beta_1 + d}},
\]
for large enough $N$,
where the infimum is taken over all estimators.
\end{thm}

{The above theorem indicates that the minimax lower bound for the density ratio estimation is the same as that for the density estimation. It is worth noting that the convergence rate provided in Theorem \ref{thm: conv-r} matches the lower bound up to a $\log(N)$ factor, meaning that the proposed estimator nearly attains the minimax bound.} 
Moreover, as will be discussed in Section \ref{sec: hd}, if the true function $r_0$ has a {low-dimensional support,} 
{then the estimation error of $\wh{r}$ estimated with the DNNs can adaptively achieve a faster convergence rate depending on the low-dimensional structure instead of the nominal dimension $d$, thus alleviating the curse of dimensionality.} 

\subsection{Conditional density estimation and multiple imputation}
\label{sec: m3}

The goal is to estimate the conditional moment $\bbm(\bX, \btheta) = \E\{ \bg(\bZ, \btheta) | \bX \}$. For a given $\btheta$, it can be simply estimated by regressing $\bg(\bZ, \btheta)$ on $\bX$. 
However, it has to be conducted repeatedly in search of the optimal solution. In this section, we propose a multiple imputation method to bypass the issue. 

To present the multiple imputation procedure, we first need to estimate the conditional density function  $p_{Y | \bX}$, 
then using it to conduct imputations {for the responses of both the source and the target samples.} 
The conditional density estimation is a conventional topic in statistics, whose development includes the kernel density estimation (\citealp{hall2004}), the nearest neighbor (\citealp{li2022-knn}), and the regression methods (\citealp{izbicki2017}). However, the existing methods are mostly restricted to certain {nonparametric forms}, such as the kernel 
or the orthogonal basis functions. We propose a new scheme for conditional density estimation which is flexible enough to accommodate a wide range of nonparametric methods. 

We note that the conditional density function is essentially a density ratio between the joint density $p_0(y, \bx)$ over the marginal density $p_0(\bx)$. However, the $\phi$-divergence based density ratio estimation method described in Section \ref{sec: m2} requires 
 the support of the denominator density covers that of the numerator density to ensure the $\phi$-divergence is well defined. For this reason, we 
express the conditional density as 
\begin{align}
    p_0(y | \bx) = \frac{p_0(y, \bx)}{p_0(\bx)} = \frac{p_0(y, \bx)}{p_0(\bx) \tilde{p}_0(y)} \tilde{p}_0(y) =: \tilde{r}_0(y, \bx) \tilde{p}_0(y),
\end{align}
where $\tilde{r}_0(y, \bx)$ is an auxiliary density ratio function between the source population $P_{\bX, Y}$ 
and an auxiliary population $\tilde{P}_{\bX, \tilde{Y}} = P_{\bX} \times \tilde{P}_Y$,
where $\tilde{P}_Y$ is supported on $\mathbb{R}$ with a known density  
 $\tilde{p}_0(y)$.  
 {Such a transformation ensures  that $\tilde{P}_{\bX, \tilde{Y}}$ is absolutely continuous with respect to the the source distribution $P_{\bX,Y}$. Hence, the $\phi$-divergence based approach for estimating the  auxillary density ratio $\tilde{r}_0(y, \bx)$ can be applied. Specifically, let $\cG_N$ be a $(d+1)$-dimensional candidate function class that satisfies Condition \ref{con: app G} below, then the density ratio $\tilde{r}_0(y, \bx)$ can be estimated via the following sample criterion } 
\begin{align}  
\wh{\tilde{r}}(y ,\bx) = \argmin_{p \in \mathcal{G}_N} \left\{  \frac{1}{n} \sumin  \ell_{1, \phi}\{p(\tilde{Y}_i, \bX_i)\}  - \frac{1}{n} \sumin  \ell_{2, \phi} \{ p(Y_i , \bX_i) \} \right\}, 
\label{eq: cde-r}
\end{align}
where $\{\tilde{Y}_i\}_{i=1}^n$  are independently sampled from $\tilde{P_Y}$ and are independent of $\{(\bX_i, Y_i) \}_{i=1}^n$. 
With $\wh{\tilde{r}}(y ,\bx)$, the conditional density is estimated by 
\be 
\wh{p}_{Y | \bX}(y | \bx) = \wh{\tilde{r}}(y ,\bx)  \tilde{p}_0(y). \label{eq:cdest} 
\ee
This facilitates  
the multiple imputation of \cite{wang-chen-09}  for the estimation of $\bbm(\bx, \btheta)$. 

{Using the  conditional density estimator $\wh{p}_{Y| \bX}(y | \bx)$,}
for any  
$\bX_i \in \{\bX_l \}_{l=1}^N$, we generate a sample $\{\tilde{Y}^{\nu}_i\}_{\nu=1}^\kappa$ independently from $\hat{p}_{Y|\bX}(y | \bX_i)$ as advocated by the multiple {nonparametric} imputation of \cite{wang-chen-09}. 
Then, the imputed moment function is 
\[
\wh{\bbm}_{\kappa}(\bX_i, \btheta) = \frac{1}{\kappa} \sum_{\nu = 1}^\kappa \bg(\bX_i,\tilde{Y}^{\nu}_i,\btheta). 
\]
{The most prominent advantage of such an imputation-based estimator is that it does not depend on any particular $\btheta$, and is in sharp contrast to the regression approach which shall regress $\bg(\bZ, \btheta)$ on each $(\bX, \btheta)$. 
} 
As in  
\cite{wang-chen-09}, it requires that $\kappa \to \infty$ as $N \to \infty$ to attain the best efficiency. To establish the convergence rate of $\wh{p}(y |\bx)$, the following conditions are required. 

\begin{con}\label{con: tilde-p}
(i) The support of $\tilde{P}_Y$ covers that of $P_Y$, and (ii) the density function of $\tilde{P}_Y$ is uniformly bounded. 
 (iii) There exist constants $B_2 > 0$ and $\beta_2 \geq 1$ such that the true conditional density function $p_{Y|\bX} \in \mathcal{H}^{\beta_2}(\mathcal{Y} \times \mathcal{X}, B_2)$. (iv) $\inf_{y \in \mathcal{Y}, \bx \in \mathcal{X}} p_{Y|\bX}(y | \bx) >  0$. 
\end{con}

\begin{con}\label{con: app G}
  The  pseudo-dimension of $\mathcal{G}_N$ satisfies 
 (i)  $\pdim(\mathcal{G}_N) \log(N) = o(N)$, and (ii) there exists a constant $c_3 > 0$ such that for large enough $n$,
 $ \inf_{p \in \mathcal{G}_N}\norm{p - p_{Y|\bX}}_{\infty} \leq c_3 \pdim(\mathcal{G}_N)^{-\frac{\beta_2}{d+1}}
$. (iii) There exists a positive constant $M_2$ such that $\norm{p}_{\infty} \leq M_2$ and $\norm{\ell_{i, \phi}''(p)}_{\infty} \leq M_2 $ for $i = 1, 2$ for every $p \in \cG_N$. 
\end{con}

\begin{con}\label{con: sub-g}
There exists a positive constant $\sigma_g > 0$ such that $\E\{\exp(\lambda \norm{\bg(\bZ, \btheta}^2 ) | \bX = \bx \} < \exp(\lambda \sigma_g^2)$ for all $0 \leq \lambda \leq \sigma_g^{-2}$ for each $\btheta \in \Theta$ and $\bx \in \mathcal{X}$. 
\end{con}

In the above conditions, Conditions \ref{con: tilde-p} (i)-(ii) are regularity conditions for the auxiliary distribution $\tilde{P}_Y$. Conditions \ref{con: tilde-p} (iii)-(iv) and Condition \ref{con: app G} are in analog to  Conditions \ref{con: r0} and \ref{con: app r} for the density ratio estimation, respectively, requiring that the true conditional density function has $\beta_2$-smoothness and the candidate function class has a sufficient approximation ability. Condition \ref{con: sub-g} assumes that $\norm{\bg(\bZ, \btheta)}$ is sub-Gaussian conditional on $\bX$. Though such a condition is not required in classic GEE literature, it is required here since we need to conduct nonparametric estimation for its conditional mean function. 

To present the result on the convergence of multiple imputation estimator $\wh{\bbm}_{\kappa}(\bX, \btheta)$, similar to \eqref{eq: error of r} of $\wh{r}$, we define the  empirical $L_2$ error of $\wh{\bbm}_{\kappa}(\bX, \btheta)$ as 
\begin{align}
&\cE_N(\wh{\bbm}_{\btheta}) =\sum_{j =1 }^r  [N^{-1}\sumiN \{\wh{m}_{\kappa j}(\bX_i, \btheta) - m_{0j}(\bX_i, \btheta)\}^2]^{1/2}, \label{eq: error of m}
\end{align}
where $\wh{m}_{\kappa j}$ and  $m_{0j}$ are the $j$-th {component} of $\wh{\bbm}_{\kappa}$ and $\bbm_0$, respectively. 

\begin{thm}\label{thm: conv rate of mi}
    Under Conditions \ref{con: dist}, \ref{con: tilde-p}--\ref{con: sub-g} and taking $\pdim(\cG_N) = O(N^{-\frac{d+1}{2\beta_2 + d+1}})$ and $\kappa \gtrsim {N}$, for any $\btheta \in \Theta$, 
    \[
   \cE_N(\wh{\bbm}_{\btheta}) \} = O_p\left(N^{- \frac{\beta_2}{2\beta_2 + d + 1}}\log^\frac{3}{2}(N) \right). 
    \]
 
\end{thm}
The proof of the above theorem is similar to that of Theorem \ref{thm: conv-r} and is given in Section B.2 of the SM.
It is known from \cite{yang1999minimax} that the $N^{- \frac{\beta_2}{2\beta_2 + d + 1}}$ rate matches the minimax lower bound for the $(d+1)$-dimensional conditional mean estimation problem. Hence,  Theorem \ref{thm: conv rate of mi} shows that 
multiple imputation estimator $\wh{\bbm}_{\kappa}(\bX, \btheta)$ have the merit of being rate optimal up to the $\log(N)$ factor, while conveniently avoiding conducting infinitely many regressions at all possible $\btheta$.
The effect of $d$ in the above rate reveals the curse of dimensionality. 
However, the accommodation of flexible uses of modern machine learning algorithms in the proposed method provides the opportunity to improve the convergence rate since the low-dimensional structure, if the underlying distribution indeed posits, can be adaptively learned by the DNNs, as will be shown in \ref{thm: hd-adapt}.

\section{Empirical likelihood inference}
\label{sec: EL}

Using the orthogonal moment function $\bPsi(\bW_i, \btheta, \wh{\bfeta})$ with $ \wh{\bfeta}(\bX_i, \btheta) = (\wh{r}(\bX_i), \wh{\bbm}_{\kappa}(\bX_i, \btheta))$ estimated by the methods in Section \ref{sec: m2} and \ref{sec: m3}, respectively,  the EL estimator of $\btheta_0$ is 
\begin{align}
    \wh{\btheta} = \argmax_{\btheta \in \Theta} L_N(\btheta)   
\end{align}
where   $L_N(\btheta)$ is the profile EL 
\begin{align}
  L_N(\btheta) =\sup\left\{ \prod_{i=1}^N p_i ~\bigg| ~ p_i \geq 0, ~\sum_{i=1}^{N} p_i = 1, ~ ~\sum_{i=1}^{N}  p_i \bPsi(\bW_{i}, \btheta,\wh{\bfeta}(\bX_i, \btheta)) = \bzero  \right\}. 
    \label{eq: EL ratio}
\end{align}

In the following Section \ref{sec: theory-el}, we will investigate the asymptotic distribution of the EL estimator $\wh{\btheta}$ and the inference for $\btheta_0$. In Section \ref{sec: hd}, we will discuss the scenario where the covariate dimension $d$ and the parameter dimension $p$ are allowed to grow the the increase of the sample size. 

\subsection{Asymptotic results for the EL inference}
\label{sec: theory-el}

In this part, we discuss the large sample properties of the EL-based estimator $\wh{\btheta}$ and then propose conﬁdence regions for $\btheta_0$ based on the EL ratio. 
To present the result, we  define $\bGamma = \E\{\partial \bPsi(\bW, \btheta_0, \bfeta_0) / \partial \btheta \}$, $\bOmega = \E\{\bPsi(\bW, \btheta_0, \bfeta_0)^{\otimes 2} \}$, and $\bSigma = (\bGamma\t \bOmega^{-1}\bGamma)^{-1}$.
The empirical estimation errors $\cE_N(\wh{r})$ and $\cE_N(\wh{\bbm}_{\btheta})$ are defined in 
\eqref{eq: error of r} and \eqref{eq: error of m}, respectively.
\begin{thm}
    \label{thm: asy dist}
    Under Conditions \ref{con: dist} and \ref{con: moment function}, if the estimation errors satisfy  
    \begin{equation}
       \mathcal{E}_N(\widehat{r})+\mathcal{E}_N(\widehat{\bbm}_{\btheta})= o_{p}(1)
        ~~\text{and}~~
        \mathcal{E}_N(\widehat{r})\mathcal{E}_N(\widehat{\bbm}_{\btheta}) = o_{p}(N^{-\frac{1}{2}}),
        \label{eq: con of eta}
    \end{equation}
    for every $\btheta \in \Theta$,
    then we have
    \begin{equation}
     \sqrt{N}(\wh{\btheta} - \btheta_0) \indist \cN\left(\bzero, \bSigma\right). 
     \label{eq: asy dist}
    \end{equation}
\end{thm}

In this theorem, the requirement for the nuisance function estimation is only via their estimation errors \eqref{eq: con of eta}. 
Specifically, under Conditions \ref{con: r0}--\ref{con: sub-g} where $r_0$ and $p_{Y|\bX}$ have the smoothness of $\beta_1$ and $\beta_2$, respectively, then \eqref{eq: con of eta} is attainable provided that 
\begin{equation}
    \frac{\beta_1}{2\beta_1 + d} + \frac{\beta_2}{2\beta_2 + d + 1} > \frac{1}{2},
\label{eq: smooth requirement}
\end{equation}
using the proposed divergence-based density ratio estimator $\wh{r}$ and the multiple-imputation estimator $\wh{\bbm}_{\kappa}$  whose convergence rates are established in Theorem \ref{thm: conv-r} and \ref{thm: conv rate of mi}. It is remarkable that the asymptotic variance of $\wh{\btheta}$ reaches the semiparametric efficiency bound {established in \cite{chen_semiparametric_2008}} for problem \eqref{eq: EE}, meaning that it has the optimal variance among the family of unbiased estimators for $\btheta_0$. 
Compared with the estimators {in \cite{chen_semiparametric_2008}}, the proposed method accommodates more flexible uses of the ML methods for the nuisance function estimation and requires milder conditions to achieve the asymptotic normality in \eqref{eq: asy dist}, as will be further discussed in Section \ref{sec: drw-theory}. 

\begin{rek}
Different from the cross-fitting adopted by \cite{chernozhukov2018} and \cite{Kallus2024} among many others in the recent literature of semiparametric inference with machine learning methods, our theoretical results do not necessarily require the sample splitting procedure. It is noted that the sample splitting may alleviate potential overfitting problems and under some conditions may lead to faster convergence of the reminder terms as shown in \cite{newey2018cross}.  However, due to the heavy computational cost of the cross-fitting, we do not consider such a procedure in this study. 
Moreover, the reduced sample size caused by the sample splitting may deteriorate the empirical performance of the ML-based estimation, especially when the original sample size is not sufficiently large.   
Detailed comparisons for the proposed whole sample and the cross-fitting methods are of future interest. 
\end{rek}

We next consider the inference for $\btheta_0$.  Let the log EL ratio be $\ell_N(\btheta) = -\log\{L_N(\btheta) / N^{-N} \}$ for every $\btheta \in \Theta$, and let $R_N(\btheta_0) = 2 \ell_N(\btheta_0) - 2\ell_N(\wh{\btheta}). 
$
The next theorem shows that the $R_N(\btheta_0)$ converges to a standard $\chi^2$ distribution. 

\begin{thm}\label{thm: wilks}
Under the same conditions as in Theorem \ref{thm: asy dist}, as $N \to \infty$, 
\[
R_N(\btheta_0) \indist  \chi^2_r.
\]
\end{thm}
The central $\chi^2$ distribution in Theorem \ref{thm: wilks}  brings convenience for 
the inference of
$\btheta$. Different from other methods such as the Wald-type inference and the GMM, we do not require the estimation of the asymptotic variance of $\wh{\btheta}$ due to the self-normalization of the EL. Theorem \ref{thm: wilks} is often referred to as Wilks's theorem, as one of the most prominent benefits of the EL-based inference for the GEEs (\citealp{qin-lawless-94}). However, with the presence of nuisance functions, the log EL ratio no longer necessarily converges weakly to a central $\chi^2$ distribution but may be a weighted sum of $\chi^2$ distributions, whose critical values require Bootstrap to approximate, as demonstrated in \cite{wang-chen-09} and \cite{hjort2009}. Due to the orthogonal estimating function, our method overcomes such a situation and restores Wilks's theorem of the log EL ratio, despite the involvement of two nuisance functions.

\subsection{Circumventing the curse of dimensionality}
\label{sec: hd}

In various modern scientific tasks, the dimension $d$ of the covariate can be very large.  In this part, we consider the inference for $\btheta$ with the presence of a high dimensional covariate.  It is known that the increase in dimensionality deteriorates the convergence rates of estimators (\citealp{stone1982optimal}). 
Recently, it has been investigated that the DNNs can adaptively approximate high-dimensional functions with low-dimensional structures (\citealp{jiao2022}). 
There have been increasing studies indicating that high-dimensional data tend to be supported on some low-dimensional manifolds in many applications, such as image analysis and natural language processing (\citealp{goodfellow2016deep}). 
Therefore, we consider the following approximate manifold support condition. 

\begin{con}[Approximate manifold support]
\label{con: manifold}

The covariate distributions  $P_{\bX}$ and $Q_{\bX}$ are concentrated on $\cM_{\rho}$, a $\rho$-neighborhood of $\cM \subset \cX$, where $\cM$ is a compact $d_{\cM}$-dimensional Riemannian manifold (\citealp{lee2006}) and 
$\cM_{\rho} = \{\bx \in \cX: \inf\{\norm{\bx - \by}_2: \by \in \cM \} \leq \rho \},~ \rho \in (0,1)$. 
\end{con}
In the above condition, the dimension $d_{\cM}$ of the manifold $\cM$ can be regarded as an intrinsic dimension of the covariate. Throughout this section, we allow the nominal dimension $d$ to diverge with the sample size, while taking the intrinsic dimension $d_{\cM}$ as a fixed constant. \cite{jiao2022} established that the fully connected DNNs can adaptively estimate a smooth function with the manifold assumption, hence alleviating the curse of dimensionality. Motivated by the development, we choose the function classes $\cF_N$ in the density ratio and $\cG_N$ in conditional density estimation as the DNNs with the ReLU activation function. The widths for $\cF_N$ and $\cG_N$ are specified as $W_1$ and $W_2$, and the depths are specified as $D_1$ and $D_2$, respectively. 
Let $\tilde{d}_{\cM} = O(d_{\cM}\log(d / \delta) / \delta^2)$ be an integer such that $d_{\cM} \leq \tilde{d}_{\cM} < d$, where $\delta \in (0,1)$ is a given constant. 
The following theorem gives the convergence rate of the DNN-based estimation of the nuisance functions under Condition \ref{con: manifold}.

\begin{thm}\label{thm: hd-adapt}
Under Conditions \ref{con: r0}--\ref{con: manifold}, let the widths and depths of $\cF_N$ and $\cG_N$ be 
\[
W_i = 114(\lfloor \beta_i \rfloor + 1)^2 \tilde{d}_{\cM}^{\lfloor \beta_i \rfloor + 1} ~~\text{and}~~D_i = 21 (\lfloor \beta_i \rfloor + 1)^2  N^{\tilde{d}_{\cM} / 2(\tilde{d}_{\cM} + 2 \beta_i ) } \lceil  \log_2(8 N^{\tilde{d}_{\cM} / 2(\tilde{d}_{\cM} + 2 \beta_i )}) \rceil, 
\]
for $i = 1$ and $2$. 
Then, the estimation errors of $\wh{r}$ and $\wh{\bbm}_{\btheta}$ satisfy 
\begin{equation}
\label{eq: conv rate manifold}
\begin{aligned}
    &\cE_N(\wh{r}) =  
    O_p\left(d^{\frac{1}{2}}  N^{-\frac{\beta_1}{\widetilde{d}_{\cM} + 2 \beta_1}} \log^{\frac{1}{2}}(N)\right)~~\text{and} \\ 
&\cE_N(\wh{\bbm}_{\btheta})  
= O_p\left(
(d+1)^{\frac{1}{2}}N^{-\frac{\beta_2} { (\widetilde{d}_{\cM} +1+ 2 \beta_2)} } \log^{\frac{3}{2}}(N)\right), \quad \hbox{respectively.} 
\end{aligned}
\end{equation} 
\end{thm}

The above theorem shows that the DNN-based estimation for the nuisance functions is adaptive to the low-dimensional manifold structure, with the convergence rates depending on the intrinsic dimension $\tilde{d}_{\cM}$ and a prefactor of the rate $\sqrt{d}$.  In comparison, the convergence rates for $\wh{r}$ and $\wh{\bbm}_{\btheta}$ established in Corollary \ref{thm: conv-r} and Theorem \ref{thm: conv rate of mi} without the manifold condition are $O_p(N^{-\frac{\beta_1}{d + 2 \beta_1}})$ and $O_p(N^{-\frac{\beta_1}{d+1 + 2 \beta_1}})$, respectively, up to some $\log(N)$ factors. Therefore, the effect of the dimensionality is substantially mitigated with the adaptivity of the DNN function classes to the underlying low-dimensional manifolds.  Compared with classic structural methods that pre-assume some low-dimensional structures, such as the additive models, the DNNs can obtain considerably improved convergence rates and circumvent the curse of dimensionality without the knowledge of the specific low-dimensional function structure.  

\begin{rek}
Aside from the manifold assumption considered above, there are several other low-dimensional structure conditions that the DNNs can be adaptive to. For example, \cite{bauer2019deep} showed if the underlying function follows the $\beta$-smooth generalized hierarchical interaction model of the order $\tilde{d}$, then the estimation error of the sigmoid-activated DNN achieves the order of $O_p(\alpha(d) N^{-\frac{\beta}{\widetilde{d}_{\cM} + 2 \beta}})$ for some $\alpha(d)$ depending on $d$. 
See also \cite{Schmidt-Hieber2020} 
for similar results.
However, the $\alpha(d)$ may depend exponentially on $d$, 
while in \eqref{eq: conv rate manifold} only the factors of the order $\sqrt{d}$ are involved. Hence, we mainly consider the manifold condition in this study. 
\end{rek}

We first discuss the inference for the fixed dimensional $\btheta$ with the presence of the covariate with a growing dimension $d$, namely $p$ is a constant while $d$ can increase with the sample size $N$. Such a scenario corresponds to the parameter depending on $Y$ but not on $\bX$. 
The following theorem specifies the regime for $(d, \tilde{d}_{\cM}, \beta_1, \beta_2, N)$, where the estimator $\wh{\btheta}$ and the log EL ratio $R_N(\btheta_0)$ have the same asymptotic distributions as those in Section \ref{sec: theory-el}. 

\begin{thm}
Under Conditions \ref{con: dist}--\ref{con: manifold} and suppose that $d = O(N^k)$ for some $k \geq 0$ and 
\begin{align}
      \frac{\beta_1}{2\beta_1 + \tilde{d}_{\cM}} + \frac{\beta_2}{2\beta_2 + \tilde{d}_{\cM} + 1}  > \frac{2+k}{4}, 
        \label{eq: hd-regime}
\end{align}
then $\sqrt{N}(\wh{\btheta} - \btheta_0) \indist \cN\left(\bzero, \bSigma \right)$ and $R_N(\btheta_0) \indist  \chi^2_r$ as $N \to \infty$. 
\end{thm}

Compared with Condition \eqref{eq: smooth requirement} under the fixed $d$ and without the manifold structure, the requirement in \eqref{eq: hd-regime} replaces the $d$ factors appeared on the denominators to the intrinsic dimension $\widetilde{d}_{\cM}$, which provides the opportunity to allow the nominal dimension $d$ grows with the polynomial rate of $N$. \cite{chen2024} also considered a growing dimension scenario and applied the shallow neural network for nuisance function estimation, where 
the dimension was allowed to increase at the rate $d = o(\sqrt{\log(N)})$. 

Next, we consider the case where both the dimensions of $\btheta$ and $\bX$ diverge, namely $p, d\to \infty$ as $N \to \infty$, which implies the number of moment restrictions $r \to \infty$ since it is no less than the number of parameters $p$ for the identification. 
The high dimensional EL without the nuisance functions has been investigated by \cite{chen-peng-qin-09}, \cite{hjort2009}, and \cite{chang-chen-chen-15}. The following extends their results to the covariate shift setting in the presence of high dimensional nuisance functions. 

\begin{thm}\label{thm: asy dist hd}
    Under Conditions \ref{con: dist}--\ref{con: manifold} and regime \eqref{eq: hd-regime}, 
    if $r^3 p^2 N^{-1} = o(1)$ and $r^3 N^{2 / \alpha - 1} = o(1)$, where 
    $\alpha > 2$ is the order of moment defined in Condition \ref{con: moment function}, then as $r, p, N \to \infty$,
    (i) for any $\bu_n \in \mathbb{R}^p$ with unit $L_2$-norm, 
\begin{align}
    &\sqrt{N} \bu_n\t \bSigma^{-1} (\wh{\btheta} - \btheta_0) \indist \mathcal{N}(0,1); 
\end{align}
(ii) the EL ratio statistic $R_N(\btheta_0)$ satisfies  
\begin{align}
    (2 r)^{-\frac{1}{2}} \{R_N(\btheta_0) - r \} \indist \mathcal{N}(0,1). 
    \label{eq: normal wilk}
\end{align}
\end{thm}
Although the estimating function involves nuisance functions, the above asymptotic distributions of $\wh{\btheta}$ and $R_N(\btheta_0)$ recover those in \cite{chang-chen-chen-15} in the absence of the nuisance functions, due to the Neyman-orthogonality of the construction for $\bPsi$. 
As established in Theorem \ref{thm: asy dist hd} (i), the normalized EL estimator $\wh{\btheta}$ remains asymptotic normal under 
$r^3 p^2 N^{-1} = o(1)$ and $r^3 N^{2 / \alpha - 1} = o(1)$. The asymptotic normality 
\eqref{eq: normal wilk} for $R_N(\btheta_0)$ is a natural substitute for the Wilks' theorem with diverging $r$, which conveniently facilitates the inference for $\btheta_0$. 

\begin{rek}
The above analyses are under the regime where $p$ and $r$ diverge at rates slower than the sample size $N$. For the ultra high dimensional cases where $p, r \gg N$, one can utilize the penalized EL  approach introduced by \cite{changNewScopePenalized2018} and \cite{changHighdimensionalEmpiricalLikelihood2021}, while imposing sparsity structures on the model parameters. 
\end{rek}

\section{Related methods}
\label{sec: relate}

In this section, we will discuss the distinctions of our proposed approach to some popular methods in the missing data and causal inference literature.

\subsection{Density ratio weighting estimation}
\label{sec: drw-theory}

In this part, we formally establish the theoretical properties of the density ratio weighting (DRW) estimation briefly discussed in Section \ref{sec: m1}, 
which is employed in \cite{chen_semiparametric_2008} and  \cite{chen2024} for the inference of GEEs
in missing data problems. 

Since the covariate shift setting $P_{Y | \bX} = Q_{Y | \bX}$ is equivalent to 
the missing at random 
condition, 
the GEE problem \eqref{eq: EE} is closely related to that considered in \cite{chen_semiparametric_2008}. By the Bayes rule, the density ratio function $r_0$ can be expressed as 
\begin{align}
r_0(\bx) = \frac{f(\bx| \delta  = 0)}{f(\bx| \delta  = 1)} = \frac{\P(\delta = 1)}{\P(\delta = 0)} \frac{\P(\delta = 0| \bX = \bx)}{1- \P(\delta = 0| \bX = \bx)}.
\label{eq: r-pi}
\end{align}
Let $\pi_0(\bx) = \P(\delta = 0| \bX = \bx)$, which is 
the propensity score function in the missing data literature. Hence, (\ref{eq: r-pi}) 
reveals that $r_0$ has a one-to-one correspondence with $\pi_0$. Therefore, the DRW estimator is essentially an IPW estimator. 
The most important advantage of the DRW estimator is that it does not need to estimate the conditional mean function $\bbm(\bX, \btheta)$. 
However, unlike most classic IPW estimators, 
where the propensity score $\pi_0$ is estimated using the logistic or the least squares regression, in the density ratio weighting for the covariate shift, we often directly estimate $r_0$ instead of $\pi_0$. Therefore, the results shown for the IPW estimators may not directly hold for the DRW estimator. In the following, 
we explore whether and when the DRW estimator is as efficient as the proposed method. 


Suppose the density ratio estimator $\wh{r}$ satisfies 
\begin{align}
        \wh{L}_N(\wh{r}) \leq \wh{L}_N(r) + O_p(\epsilon_N^2), ~~\text{for all } r\in \cF_N, 
\label{eq: L-obj}
\end{align}
where  $ \cF_N$ is the function class where the density ratio estimator is chosen from,  $\epsilon_N$ is a positive sequence satisfying $\epsilon_N = o(N^{-\frac{1}{2}})$, and the objective function $ \wh{L}_N(r)$ is defined as 
\begin{align}
     \wh{L}_N(r) = \frac{1}{n} \sumin  \ell_1(\bX_i, r(\bX_i))-\frac{1}{m}\sum_{i = n+1}^{n+m} \ell_2(\bX_i, r(\bX_i)), \nn 
\end{align}
where $\ell_i (i = 1,2)$ are  
not necessarily the $\ell_{i, \phi} (i = 1,2)$ introduced in \eqref{eq: dre}. 

With an estimated $\widehat{r}(\bx)$, the DRW moment function for the source sample  is 
\begin{equation*}
    {\bg}\drw(\bZ_{i}, \btheta,\widehat{r}) =
    \widehat{r}(\bX_{i}) \bg(\bZ_{i}, \btheta)  ~~~\text{for} ~i = 1,\dots, n,
\end{equation*}
from which we can obtain an estimator of $\btheta_0$ defined as 
\begin{align}
    \wh{\btheta}\drw = \argmax_{\btheta \in \Theta} L_{n}\drw (\btheta, \wh{r}), 
    \label{eq: drw-est}
\end{align}
where $L_{n}\drw (\btheta,\wh{r})$ is the profile EL ratio 
\begin{align}
    L_n\drw(\btheta,\wh{r}) = \sup\left\{ \prod_{i=1}^n  p_i ~\bigg| ~ p_i \geq 0, \sum_{i=1}^n p_i = 1,  \sum_{i=1}^n p_i {\bg}\drw(\bZ_{i}, \btheta,\widehat{r}) = \bzero  \right\}. \nn 
\end{align}

For a given $\ell_2(\bx, r)$, we let $\bbm_{\ell}(\bx) = \bbm(\bx, \btheta_0) \{ \partial \ell_{2}(\bx, r_0) / \partial r \}^{-1}$. The following conditions are required to establish asymptotic properties of $\wh{\btheta}\drw$. 

\begin{con}\label{con: l1-l2}
(i) The objectives $\ell_1(\bx, r)$ and $\ell_2(\bx, r)$ are three-times continuously differentiable with respect to $r$ and  $\inf_{\bx \in \cX} \partial \ell_{i}(\bx, r_0) / \partial r > 0$ for $i = 1, 2$. (ii) The partial derivatives satisfy $\partial \ell_1(\bx, r)/ \partial r = r(\bx) \partial \ell_2(\bx, r)/ \partial r$.
\end{con}

\begin{con}\label{con: hat-r}
(i) The estimation error of $\wh{r}$ satisfies $\norm{\wh{r} - r_0}_{L_2(P)} = O_p(\delta_N)$ for some $\delta_n = o(N^{-\frac{1}{4}})$. 
(ii) The bracketing integral (see \citealp{van2000asymptotic}) of $\mathcal{F}_N$ satisfies $
J_{[\ ]}(\delta_N, \cF_N, L_2(P)) = o(1)$. 
(iii) For every $i = 1, \cdots p$, there exists some $\tilde{m}_{\ell,j} \in \mathcal{F}_N$ such that $\norm{m_{\ell, j}(\bX) - \tilde{m}_{\ell,j}(\bX)}_{L_2(P)} = o(N^{-\frac{1}{4}})$, where $m_{\ell, j}(\bx)$ is the $j$-th element of $\bbm_\ell(\bx)$. 
\end{con}

Condition \ref{con: l1-l2} is regarding the requirements of the objective function and ensures $r_0$ is the solution to the population objective function.   
Condition \ref{con: hat-r} collects the assumptions for $\wh{r}$ and the function class $\cF_N$ to which $\wh{r}$ belongs. Specifically, Condition \ref{con: hat-r} (i) requires the $L_2$-estimation error of $\wh{r}$ to be $o_p(N^{-\frac{1}{4}})$. In comparison, our result only requires the \textit{product} of the estimation errors of $\wh{r}$ and $\wh{\bbm}_{\btheta}$ to be 
$o_p(N^{-\frac{1}{2}})$, achieving more robustness against the nuisance function estimation errors.  Condition \ref{con: hat-r} (ii) is a restriction for the complexity of $\mathcal{F}_N$, which is needed for stochastic equicontinuity. Condition \ref{con: hat-r} (iii) assumes that each element of $\bbm_\ell(\bx)$ can be approximated sufficiently well by the function class $\cF_N$. 
It is worth noting that such a condition implicitly brings more smoothness 
for the conditional mean function $\bbm(\bx, \btheta_0)$.    
With the above conditions, the asymptotic distributions of the DRW estimator $\wh{\btheta}\drw$ and the associated log EL ratio statistics $ {R}\drw_n(\btheta_0)$ can be derived. 
\begin{thm}\label{thm: drw}
    Under Conditions \ref{con: dist}--\ref{con: r0}, \ref{con: app r}. (iii),  \ref{con: l1-l2} , and \ref{con: hat-r}, the DRW estimator $\wh{\btheta}\drw$ satisfies 
\begin{align}
    \sqrt{N}(\wh{\btheta}\drw - \btheta_0) \indist \bU \sim \mathcal{N}\left( 0, 
   \bSigma
    \right),
\end{align}
 and the log EL ratio statistics 
${R}\drw_n(\btheta_0) \indist \bU\t [\E\{\bg(\bZ, \btheta_0)^{\otimes 2}\} ]^{-1}\bU$. 
\end{thm}
The theorem reveals that the DRW estimator $\wh{\btheta}\drw$ attains the semiparametric efficiency bound as the proposed estimator. However, it requires more stringent conditions on both the density ratio estimation error and the approximation ability of $\cF_N$, which are not necessarily needed by our method. More importantly,  the limiting distribution of the log EL ratio ${R}\drw_n(\btheta_0)$ has a weighted $\chi^2$ limiting distribution, since the covariance of $\bU$ does not match with $\E\{\bg(\bZ, \btheta_0)^{\otimes 2}\}$. Consequently, for the inference of $\btheta_0$, the above density ratio weighting method requires a Bootstrap procedure (\citealp{chen2024}), which brings considerable computation burden especially when $\wh{r}$ is estimated with complex ML algorithms such as the DNNs, while our proposed method can conveniently employ the Wilk's theorem of the EL ratio statistics.

\subsection{Double machine learning methods}

Our work is also closely related to the classical semiparametric estimation literature on constructing asymptotic normal estimators for low dimensional parameters with the presence of infinitely dimensional nuisance functions. 

Building upon the Neyman orthogonality condition, our modified moment function shares similar spirits as the class of doubly robust estimators (\citealp{robins_estimation_1994} and \citealp{rotnitzky2012improved}) and recently proposed double machine learning methods (\citealp{chernozhukov2018}). However, this study has the following important distinctions. First, both the doubly robust and the double machine learning literature commonly deal with linear functional estimation, such as the average treatment effect. Under such cases, the nuisances are typically the propensity score function $\pi_{0}(\bX) = \P(\delta = 0 | \bX)$  and the conditional mean function $m(\bX) = \E(Y| \bX)$, which both can be easily estimated by solving a regression problem. However, our interested GEE problem is more challenging, due to the presence of the nuisance function $\bbm(\bX, \btheta) = \E\{\bg(\bZ, \btheta) | \bX \}$, which requires estimating for all possible $\btheta$. Therefore, our work complements the line of research of the doubly robust and double machine learning methods, by providing an effective approach to handle such parameter-dependent nuisance function. Utilizing the idea of the multiple imputation, we circumvent directly estimate $\bbm(\bX, \btheta)$ at infinitely many $\btheta$ but only requires the estimation of the conditional density function $p(y | \bX)$. Instead of the conventional kernel smoothings, our novel method for the estimation $p(y | \bX)$ 
can employ a broad array of machine-learning algorithms.  Moreover, the sample splitting procedure required in the DML can be bypassed in this study as discussed in Remark 1. 

\section{Simulation Study}
\label{sec: sim}

This section reports the simulation results for the proposed methods, including the density ratio estimation, the conditional density estimation, and the inference for the GEEs. 

\subsection{Numerical results of density ratio estimation}
\label{sec: sim-dre}

In this part, we carried out simulations to evaluate the performances of the proposed density ratio estimation and compared it with other popular density ratio estimation methods.

The covariate of source sample $\{ \bX_{i} \}_{i = 1}^n$  
and the target sample $\{ \bX_{n+i} \}_{i = 1}^{n+m}$ were generated as independent copies of $\bX^0 = (X_{1}^0,\dots, X_{d}^0)\t$ and $\bX^1 = (X_{1}^1,\dots, X_{d}^1)\t$, respectively.
The sample sizes were chosen within the range $n \in \{ 1000, 2000, 5000\}$ for the source sample and $m = n / 2$ for the target sample, to accommodate the common case where the source sample usually has more observations than the target sample.
We considered two settings for the distributions of $\bX^0$ and $\bX^1$, corresponding to the compact and uncompact supports, respectively. In Setting S1, $\{X_i^0\}_{i=1}^d$ were independent distributed as Uniform$(0,1)$, and $\{X_i^1\}_{i=1}^d$ were independent distributed as Beta$(6/5, 6/5)$. In Setting S2 $\bX^0$ was distributed as $\mathcal{N}(\bzero, \bm{I}_d)$, where $\bm{I}_d$ is the $d$-dimensional identity matrix,  and $\bX^1$ was distributed as $\mathcal{N}(\bzero, \bm{\Sigma}_d)$, where $\bm{\Sigma}_d = (\sigma_{i,j})_{d\times d}$ for $\sigma_{i,j} = 0.5^{|i-j|}$. 
The dimensions were chosen as $d = 5$ and $20$, respectively, to evaluate the estimation performances as the increase of the dimension. 

For the proposed divergence-based density ratio (DDR) estimation, we chose the $\phi$-divergence in \eqref{eq: D phi} as the KL-divergence and estimated $r_{0}$ by 
\[
\widehat{r} = \argmax_{r \in \mathcal{F}_{N}}  \left\{ \frac{1}{m} \sum_{i=n+1}^{n+m} \log \left( r(\bX_{i}) \right) - \frac{1}{n}\sum_{i=1}^{n} r(\bX_{i})  \right\}. 
\] 
The function class $\cF_N$ was chosen as the deep neural network, 
whose tuning parameters were selected by the three-fold cross-validation. We chose the ReLU function as the activation function and adopted the Adam as the optimization algorithm. 
For comparison, we also estimated the density ratio function with three commonly used methods: (1) the kernel mean matching (KMM) proposed by \cite{gretton2009}; 
(2) the kernel smoothing (KS) method that first obtains the kernel smoothing density estimates $\hat{p}(\bx)$ and $\hat{q}(\bx)$ for the source and target distributions, {where the bandwidths were selected from the leave-one-out cross validation}, then taking their ratio; 
(3) the probabilistic classiﬁcation (PC) approach adopted by \cite{Lei2021}, which used the ratio of posterior classification probabilities to estimate the density ratio function. 
Simulation results were based on 300 repetitions. 
The performances of the estimators were evaluated by the mean squared error (MSE) calculated as $n^{-1}\sum_{i=1}^n \{  \widehat{r}(\bX_{i}) - r_{0}(\bX_{i})\}^2$.  

\begin{table}[ht]
\scriptsize  
    \centering
    \caption{Empirical average of mean squared errors (MSE) of $\widehat{r}$  based on $300$ repetitions of the proposed divergence-based density ratio (DDR) estimation, the kernel mean matching (KMM), the kernel smoothing (KS), and the probabilistic classification (PC) methods. The empirical standard deviations of the MSEs 
    are in parentheses. }
    \begin{tabular}{c  c  c c c c c c c c}

      \hline\hline
       \multirow{2}{*}{Setting} & \multirow{2}{*}{$n$} & \multicolumn{4}{c}{$d = 5$} & \multicolumn{4}{c}{$d = 20$} \\
       \cmidrule(lr){3-6}  \cmidrule(lr){7-10} 
    & & DDR & KKM   & KS & PC   & DDR & KKM   & KS & PC  \\ 
      \hline
      \multirow{6}{*}{S1} &  \multirow{2}{*}{$1000$}& 
      0.64 & 0.75 & 0.86 & 0.81 & 0.93 &  1.24 & 1.85 & 1.56\\ 
     &  & (0.35) & (0.29) & (0.31) & (0.28) & (0.43) & (0.52) & (0.68) & (0.59)  \\ 
       \cmidrule(lr){3-6}  \cmidrule(lr){7-10} 
      &  \multirow{2}{*}{$2000$}& 
      0.34 & 0.44 & 0.52 & 0.48 & 0.58 &  0.71 & 1.16 & 0.92   \\ 
     &  & (0.19) & (0.15) & (0.20) & (0.16) & (0.30) & (0.25) & (0.36) & (0.24)  \\ 
       \cmidrule(lr){3-6}  \cmidrule(lr){7-10} 
      &  \multirow{2}{*}{$5000$}& 
      0.22 & 0.31 & 0.38 & 0.34 & 0.36 &  0.45 & 0.81 & 0.67   \\ 
     &  & (0.09) & (0.11) & (0.09) & (0.07) & (0.13) & (0.15) & (0.25) & (0.19)  \\ 
      \hline
        \multirow{6}{*}{S2} &  \multirow{2}{*}{$1000$}& 
      0.74 & 0.82 & 0.93 & 0.85 & 1.15 & 1.38 & 1.94 & 1.73 \\ 
     &  & (0.36) & (0.25) & (0.30) & (0.36) & (0.50) & (0.62) & (0.71) & (0.66)  \\ 
       \cmidrule(lr){3-6}  \cmidrule(lr){7-10} 
      &  \multirow{2}{*}{$2000$}& 
      0.37 & 0.51 & 0.65 & 0.49 & 0.64 &  0.78 & 1.25 & 1.04   \\ 
     &  & (0.16) & (0.13) & (0.22) & (0.15) & (0.34) & (0.30) & (0.41) & (0.28)  \\ 
       \cmidrule(lr){3-6}  \cmidrule(lr){7-10} 
      &  \multirow{2}{*}{$5000$}& 
      0.24 & 0.37 & 0.44 & 0.40 & 0.43 &  0.52 & 0.93 & 0.79   \\ 
     &  & (0.08) & (0.09) & (0.13) & (0.12) & (0.20) & (0.19) & (0.28) & (0.17)  \\ 
      \hline \hline
    \end{tabular}\label{tab: sim-dr}
\end{table}

As indicated in Table \ref{tab: sim-dr}, the proposed DDR for the density ratio estimation achieved the best finite sample performances among the two settings. 
The KMM using the reproducing kernel Hilbert space had the second smallest MSE in most cases, followed by the PC method that used the posterior classification probability to indirectly estimate the density ratio. The conventional kernel smoothing (KS) method had the worst performances among the four candidates. 
The DDR had significantly improved estimation accuracy over the other methods especially for $d = 20$, suggesting the advantage of the proposed method for the large dimensional scenarios. On the other hand, it was noted that the standard deviations of the DDR estimates were relatively large in some cases with small and moderate sample sizes, which became smaller as the sample size were larger. 

\subsection{Numerical results of conditional density estimation}
\label{sec: sim-cde}

We also conducted simulations to evaluate the finite sample performance of the proposed conditional density estimation methods in Section \ref{sec: m3}. The estimation was conducted using the source sample $\source$, where the covariates $\{ \bX_{i} \}_{i = 1}^n$ generated independently from the Uniform$(0,1)^d$ distribution, and the responses $\{Y_i\}_{i = 1}^n$ were generated according to the following three models: 
\begin{align}
&\text{(M1):} ~~~~~ Y_i = 0.5 \textstyle \sum_{k = 1}^{\lfloor d/2 \rfloor}\sum_{j = 2k} X_{j,i} - 0.5  \sum_{k = 1}^{\lfloor d/2 \rfloor} \sum_{j = 2k-1} X_{j,i} + \epsilon_i,\\ 
&\text{(M2):} ~~~~~ Y_i =  \sin\left(\pi \textstyle \sum_{k = 1}^{\lfloor d/2 \rfloor} \sum_{j = 2k}X_{j,i}\right) + \epsilon_i, \nn \\ 
&\text{(M3):} ~~~~~ Y_i = \mathds{1} \left( \textstyle \sum_{k = 1}^{\lfloor d/2 \rfloor} \sum_{j = 2k}X_{j,i} < \sum_{k = 1}^{\lfloor d/2 \rfloor} \sum_{j = 2k-1} X_{j,i}\right) + \epsilon_i,
\end{align}
where the regression functions are linear, trigonometric, and piecewise constant, respectively, and the noises  $\{\epsilon_i\}_{i = 1}^n$ were independently distributed as $\mathcal{N}(0,\sigma^2_X)$, where $\sigma^2_X  = \max(0.5, |X_{i,1}|)$. 
The dimensions were chosen as $d = 5$ and $20$, respectively. 
For the proposed ratio transformed 
conditional density estimation (RTCDE),
we chose the auxiliary distribution $\tilde{P}_{Y}$ as the standard normal distribution to generate $\{\tilde{Y}_i\}_{i=1}^n$ in \eqref{eq: cde-r}.
The candidate function class $\mathcal{G}_N$ was chosen as the neural network, whose width and depth were selected from five-fold cross-validations. For comparison, we also conducted the conditional kernel density estimation (KCDE), 
the least squares conditional density estimation (LSCDE, \citealp{sugiyama2010least}) that uses the RKHS as its function class, and the FlexCode (\citealp{izbicki2017}) method, which reformulates the conditional density estimation as a non-parametric orthogonal series problem. 
The sample sizes were chosen as $n \in \{ 1000, 2000, 5000\}$. To measure the accuracy of the estimates, we computed the empirical MSE $n^{-1}\sumin \{\hat{p}(Y_i| \bX_i) - p_{Y|\bX}(Y_i, \bX_i) \}^2$. 

\begin{table}[ht]
\scriptsize  
    \centering
    \caption{Empirical average of mean squared errors (MSE) of the estimated conditional density based on $300$ repetitions of the proposed ratio transformed conditional density estimation (RTCDE), the KCDE, the LSCDE, and the FlexCode. The empirical standard deviations of the MSEs of the $300$ repetitions for each method are reported in parentheses. }
    \begin{tabular}{c  c  c c c c c c c c}
      \hline\hline
       \multirow{2}{*}{Models} & \multirow{2}{*}{$n$} & \multicolumn{4}{c}{$d = 5$} & \multicolumn{4}{c}{$d = 20$} \\
       \cmidrule(lr){3-6}  \cmidrule(lr){7-10} 
    & & RTCDE & KCDE   & LSCDE & FlexCode   &RTCDE & KCDE   & LSCDE & FlexCode \\ 
      \hline
      \multirow{6}{*}{M1} &  \multirow{2}{*}{$1000$}& 
      0.34 & 0.41 & 0.30 & 0.37 & 0.58 & 0.98 & 0.64 & 0.72 \\ 
     &  & (0.13) & (0.17) & (0.15) & (0.11) & (0.30) & (0.26) & (0.29) & (0.33)  \\ 
       \cmidrule(lr){3-6}  \cmidrule(lr){7-10} 
      &  \multirow{2}{*}{$2000$}& 
      0.13 & 0.23 & 0.14 & 0.17 & 0.32 & 0.53 & 0.36 & 0.41   \\ 
     &  & (0.08) & (0.09) & (0.09) & (0.06) & (0.12) & (0.16) & (0.14) & (0.19)  \\ 
       \cmidrule(lr){3-6}  \cmidrule(lr){7-10} 
      &  \multirow{2}{*}{$5000$}& 
      0.09 & 0.17 & 0.09 & 0.10 & 0.22 &  0.37 & 0.28 & 0.30   \\ 
     &  & (0.05) & (0.09) & (0.04) & (0.04) & (0.13) & (0.15) & (0.11) & (0.13)  \\ 
      \hline
        \multirow{6}{*}{M2} &  \multirow{2}{*}{$1000$}& 
      0.53 & 0.89 & 0.71 & 0.49 & 0.92 & 1.51 & 1.07 & 0.86 \\ 
     &  & (0.23) & (0.30) & (0.26) & (0.21) & (0.34) & (0.50) & (0.29) & (0.28) \\ 
       \cmidrule(lr){3-6}  \cmidrule(lr){7-10} 
      &  \multirow{2}{*}{$2000$}& 
      0.25 & 0.46 & 0.30 & 0.24 & 0.43 &  0.83 & 0.52 & 0.45   \\ 
     &  & (0.12) & (0.18) & (0.11) & (0.09) & (0.20) & (0.24) & (0.18) & (0.20) \\ 
       \cmidrule(lr){3-6}  \cmidrule(lr){7-10} 
      &  \multirow{2}{*}{$5000$}& 
      0.16 & 0.25 & 0.18 & 0.16 & 0.28 &  0.53 & 0.34 & 0.31   \\ 
     &  & (0.09) & (0.11) & (0.06) & (0.05) & (0.16) & (0.19) & (0.11) & (0.13)  \\ 
        \hline
        \multirow{6}{*}{M3} &  \multirow{2}{*}{$1000$}&  
      0.67 & 0.91 & 0.86 & 0.77 & 1.18 & 1.55 & 1.34 & 1.27 \\ 
     &  & (0.18) & (0.36) & (0.31) & (0.25) & (0.53) & (0.65) & (0.42) & (0.50)  \\ 
       \cmidrule(lr){3-6}  \cmidrule(lr){7-10} 
      &  \multirow{2}{*}{$2000$}& 
      0.31 & 0.59 & 0.44 & 0.40 & 0.52 &  0.79 &  0.66 & 0.59   \\ 
     &  & (0.12) & (0.18) & (0.13) & (0.15) & (0.23) & (0.27) & (0.19) & (0.26)  \\ 
       \cmidrule(lr){3-6}  \cmidrule(lr){7-10} 
      &  \multirow{2}{*}{$5000$}& 
      0.18 & 0.31 & 0.25 & 0.21 & 0.27 & 0.43 & 0.37 & 0.34   \\ 
     &  & (0.07) & (0.11) & (0.09) & (0.07) & (0.11) & (0.18) & (0.13) & (0.12)  \\ 
      \hline \hline
    \end{tabular}\label{tab: sim-cde}
\end{table}

Table \ref{tab: sim-cde} suggests that the proposed RTCDE outperformed the other three methods in most experiments. 
For the small sample ($n = 1000$) and low dimensional ($d = 5$) scenarios, the LSCDE and the FlexCode had slightly smaller MSE than the RTCDE for the linear model (M1) and trigonometric model (M2), while the RTCDE showed faster convergence rates and had the best performances in large samples. The MSEs of the four estimators under the M3 model were larger than those under M1 and M2 because the underlying regression function was discontinuous, where the RTCDE still had superior performances compared to the others under such challenging cases. 

\subsection{Numerical results of estimation and inference of the GEE}
\label{sec: sim-gee}

We now present simulation results that examine the estimation accuracy of the estimator $\wh{\btheta}$ based on the orthogonal estimating functions and the empirical coverage  of the proposed inference procedure.

For the experiment results presented below, the covariates for the source and the target samples were generated in the same way as Setting S1 in Section \ref{sec: sim-dre}, where the dimension was $d = 5$,  and the responses were generated according to Model M2 in Section \ref{sec: sim-cde}. 
The target parameter was $\theta_0 = Q_{Y}^{-1}(1/2)$, namely the median of $Y$ for the target distribution, since it corresponds to a nonlinear estimating equation where the conventional AIPW cannot apply. In Section F of the SM, we report additional simulation results for the setting with $d = 20$, and  results for the inference of the mean of $Y$ for the target distribution.

The methods for comparison included the density ratio weighting (DRW), which is equivalent to the IPW of \cite{chen2024}, the multiple imputations (MI) proposed by \cite{wang-chen-09}, the proposed method having both the density ratio weighting and the multiple imputations with estimated nuisance functions (DRW-MI-E), the localized debiased machine learning (LDML)  introduced by \cite{Kallus2024}, and the covariance balancing (\citealp{imai2014}). The nuisance functions in the first four methods were all estimated with the deep neural networks for comparison fairness, whose widths and depths were chosen by the three-fold cross-validation. When using the multiple imputations, $\kappa = N/2$ imputations were made for each observation point.  
In addition, to evaluate the effects of nuisance function estimation errors, we also considered an oracle version of the DRW-MI, where the density ratio and the conditional density functions were used as the true ones (DRW-MI-T). To obtain the $95\%$ confidence intervals, we employed the Wilks theorem for the DRW-MI-E and DRW-MI-T methods and the bootstrap approximations for the others, since they do not admit the Wilk theorem. Alternatively, one can use the asymptotic normality of the estimators of these four methods, where the asymptotic variances should be estimated.

Table \ref{tab: sim-quantile} reports the performances of the six methods based on $300$ simulation replications, where the estimation performances were measured with the empirical bias, standard deviation, and the MSE, and the inference performances were reflected from the empirical coverage probability and length of the confidence intervals (CI).   
The proposed DRW-MI-E method improved both the estimation and coverage performances than the DRW and MI alone, with the MSE converging to $0$ and the coverage probability approaching the nominal level of 0.95.  It is worth noting that the simulation results of the DRW-MI-E were comparable to that of the oracle method DRW-MI-T, where the nuisance functions used the true values, confirming the theoretical analysis that estimation for the nuisances does not have the first-order effect on the proposed orthogonal estimating function. The LDML also employed the same form of estimation function, while resorting to a two-step method for the estimation of the conditional mean function, which depended on an initial estimator and required three-fold sample splitting. As a result, its empirical performances were not as competitive as the DRW-MI-E that used the MI for the conditional mean function estimation. The CB method had the worst estimation accuracy and under-coveraged confidence intervals, since its theories require correctly specified balancing functions, 
which could not be satisfied under the nonlinear response models of the simulations.

\begin{table}[ht]
\scriptsize  
    \centering
    \caption{\scriptsize  
 Empirical estimation and inference results for the median of the target population, based on $300$ simulation replications. The five methods considered are the density ratio weighting (DRW), the multiple imputations (MI), the proposed method with both the density ratio weighting and the multiple imputations using the estimated nuisance functions (DRW-MI-E), the DRW-MI using the true nuisance functions (DRW-MI-T), the localized double machine learning (LDML), and the covariance balancing (CB). The nominal coverage probability of the confidence interval is 0.95.   }
    \begin{tabular}{c  c  c c c c c }

      \hline\hline
   & Methods & Bias & Std.dev & MSE & Coverage  & Length of CI \\ 
\hline 
\multirow{6}{*}{$n = 1000$} & DRW & -0.0282 & 0.1759 & 0.0304 & 0.8791 & 0.7216  \\ 
& MI  & -0.0256  & 0.1801 & 0.0331 & 0.8900 & 0.7404 \\  
& DRW-MI-E & -0.0221 & 0.1648 & 0.0275 &  0.9326 & 0.7719 \\ 
& DRW-MI-T & -0.0194 & 0.1610 & 0.0266 &  0.9437 & 0.7805 \\
& LDML & 0.0239 & 0.1810 & 0.0333 & 0.9048 & 0.7914 \\ 
& CB & -0.0508 & 0.1964 & 0.0395 &  0.6612 &  0.8382 \\ 
\hline 
  \multirow{6}{*}{$n = 2000$}& DRW & 0.0204 & 0.1236  & 0.0157 & 0.8920 & 0.5083\\ 
& MI  & -0.0216 & 0.1217 & 0.0153 &  0.9138 &  0.4885 \\  
& DRW-MI-E & -0.0180 & 0.1139 & 0.0130 & 0.9422 & 0.4729\\ 
& DRW-MI-T & 0.0153 & 0.1062 & 0.0112 & 0.9540 & 0.4693\\ 
& LDML & 0.0211 & 0.1302 & 0.0173 & 0.8987 & 0.4910 \\ 
& CB & -0.0452 & 0.1516 & 0.0236 & 0.6865 & 0.5283\\ 
\hline 
\multirow{6}{*}{$n = 5000$} & DRW & 0.0159  & 0.0839  & 0.0073  & 0.9104 & 0.3665 \\ 
& MI  & -0.0164  & 0.0801 & 0.0067  & 0.9312 &  0.3572\\  
& DRW-MI-E & -0.0135 & 0.0745 & 0.0058 & 0.9562  & 0.3691 \\ 
& DRW-MI-T & -0.0129 & 0.0716 & 0.0052 & 0.9524  & 0.3634 \\ 
& LDML &  0.0160  & 0.0848 & 0.0074 & 0.9215 & 0.3820\\ 
& CB & 0.0312 & 0.1209 & 0.0151 & 0.6928 & 0.4343 \\ 
\hline \hline
    \end{tabular}
\label{tab: sim-quantile}
\end{table}

\section{Case Study}
\label{sec: case}

Ground level ozone (O$_3$), as an air pollutant, has been at an elevated level across North China (\citealp{li2019anthropogenic}) in the last decade. In this section, we demonstrate that the proposed method is well-suited for the transfer learning of the inference for the O$_3$ levels. 

We focus on four major cities in North China, including Beijing, Xian, Jinan, and Taiyuan, where we used the first three cities as the source domain and Taiyuan as the target domain. The study period was the spring (March 1 to May 31) of 2018, a season when the ozone level is generally high. The response variable was the hourly O$_3$ obtained from China Meteorological Administration (CMA) monitoring sites, and the covariates included hourly PM$_{2.5}$, PM$_{10}$,  nitrogen dioxide (NO$_{2}$), surface total solar radiation (TSR), surface air temperature (TEMP), relative humidity (HUMI), boundary layer height (BLH),  the low (LCC), medium (MCC) and high (HCC) cloud cover percentages, as well as the one to three hour lagged terms of the above variables, where the first three variables were obtained from China Environmental Monitor Center 
sites in each city, the TSR data was collected from CMA,  and the others were from the European Center for Medium-Range Weather Forecasts (ECMWF). We also included a DAY variable that counts for the 
number of days since March 1st to reflect the increasing radiation in the spring, which is highly statistically significant in modeling the O$_3$ as shown in \cite{li2021radiative}. Our goal was to utilize 
the O$_3$ observations and the covariates of the source sample to assist the inferences at the target population of the O$_3$ in Taiyuan. 
To investigate the performances of the transfer learning methods, we assumed only the covariate variables of the target domain Taiyuan were observable during their implementations, while the true O$_3$ levels of the target sample were used to evaluate the quality of the transfer learning.

Distinctions between the distributions of some key variables of the target and the source samples are illustrated in Figure 1 of the SM, which reveals that directly using the source samples to make inferences about the O$_3$ of the target population would introduce biases. To apply the proposed method, we first estimated the covariate density ratio of the two samples and used the source sample to estimate the conditional density function. The density ratio and the conditional density 
functions were estimated with the neural networks, whose widths and heights were chosen from the five-fold cross-validation. From the estimated conditional density function, we conducted the multiple imputations with the number of imputations $\kappa = 200$. Figure \ref{fig: impute-O3} shows the $2.5\%$ and $97.5\%$ empirical quantiles as well as the empirical mean of the imputed values for O$_3$ of the target sample. The figure shows that most of the true O$_3$ values of the target were located within the $95\%$ prediction region obtained from the multiple imputations, and the mean of the imputed values were well approximated to the true ones. Such a result not only verifies that the conditional density of the target sample was similar to that of the source, but also shows that our multiple imputation method produced high-quality surrogates for the  O$_3$ on the target domain. 
\begin{figure}[h]
    \centering
    \caption{
\scriptsize  
    Illustration for the results of the multiple imputations for O$_3$ the target sample. The upper and lower boundaries of the blue region are the $2.5\%$ and $97.5\%$ empirical quantiles of the $200$ imputations. The blue dotted line is the empirical mean of the imputed values. The red line indicates the true O$_3$ levels of the target sample.}
    \includegraphics[width = 0.65\textwidth]{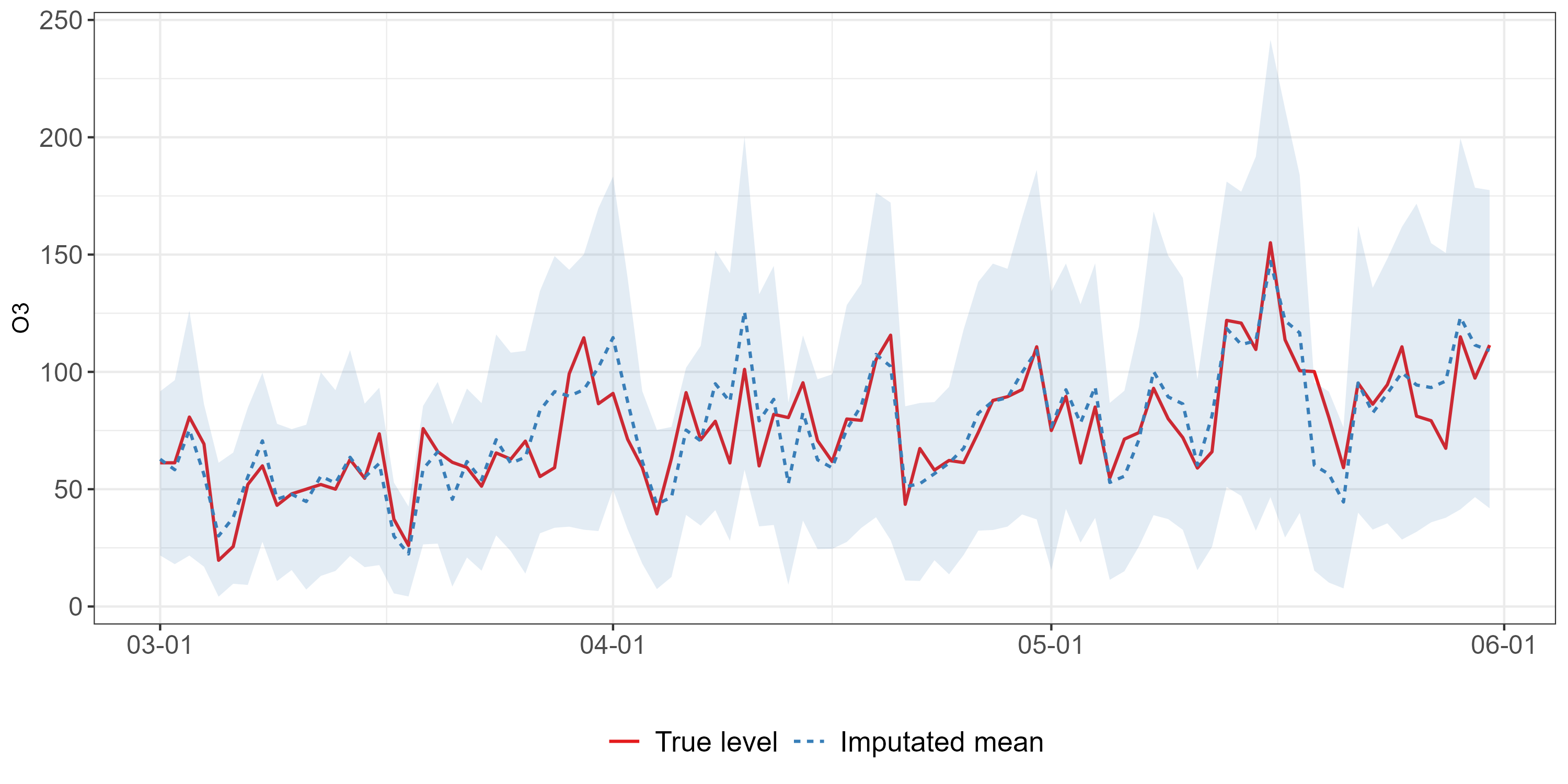}
    \label{fig: impute-O3}
\end{figure}

We considered the estimation and inference for the mean and the $\alpha$-quantiles ($\alpha = 25\%, 50\%$, and $75\%$) of the O$_3$ of the target domain in Taiyuan.  The methods include the multiple imputation (MI), the density ratio weighting (DRW), and the proposed method 
(DRW-MI). {Since the first two methods do not have Wilks' theorem, their confidence intervals (CIs) were derived by Bootstraps, where the CIs were derived based on the empirical $2.5\%$ and $97.5\%$ quantiles of the estimates from $200$ resamplings}.
The CIs for the DRW-MI were via Wilks' theorem.  As a baseline, we also considered an oracle method that used the O$_3$ of the target sample to conduct the inference for the four estimands, {including the mean and the three quantiles of the O$_3$ of the target domain,}
where their CIs were obtained based on the asymptotic normalities. As reported in Figure \ref{fig: O3-inference}, the CIs of the MI method were the largest among the four methods, indicating its being less favorable in terms of statistical efficiency. The estimation based on the DRW had a high proportion of overestimates compared with the estimation using the target sample.  The proposed DRW-MI achieved the lowest bias among the three transfer learning methods, which verifies 
that the estimation based on the orthogonal estimating functions that used both the DRW and the MI can be regarded as a one-step bias correction of the DRW estimation. 
Moreover, the CIs of the DRW-MI had shorter lengths than those obtained with the target O$_3$ observations, indicating the benefit of the TL since it utilized both the information of the source and the target sample. 

\begin{figure}[h]
\caption{
\scriptsize  
Estimation and $95\%$ confidence intervals for the mean and three quantiles of the O$_3$ of the target population obtained from the target sample, the multiple imputations (MI), the density ratio weighting (DRW), and the density ratio weighting with multiple imputations (DRW-MI), respectively. As a comparison baseline, the red dotted line indicates the estimated value of the O$_3$ with the target sample. }
    \centering
    \begin{subfigure}[t]{0.24\textwidth}
    \centering
    \caption{Mean}
    \includegraphics[width=0.9\textwidth]{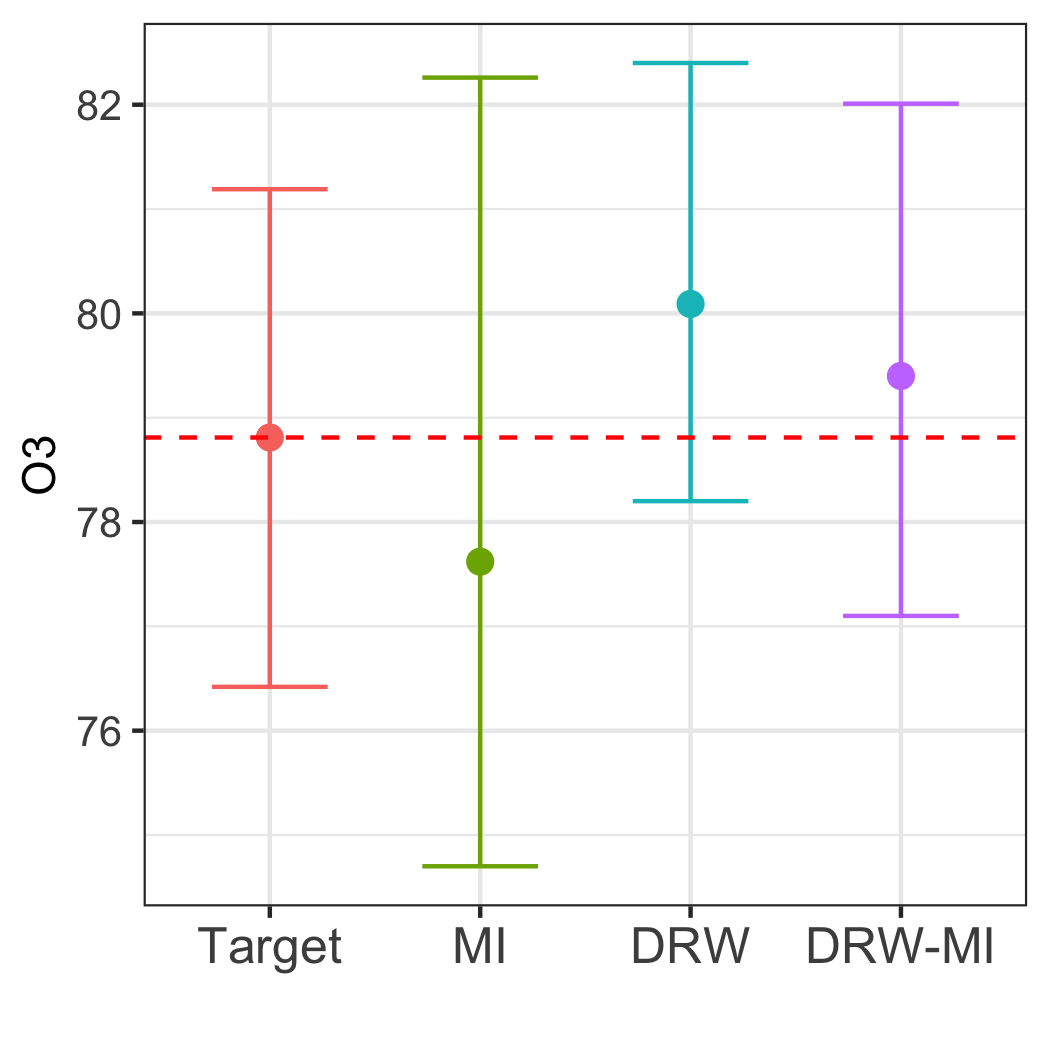}
\end{subfigure}
\begin{subfigure}[t]{0.24\textwidth}
    \centering
        \caption{$25\%$-quantile}
    \includegraphics[width=0.9\textwidth]{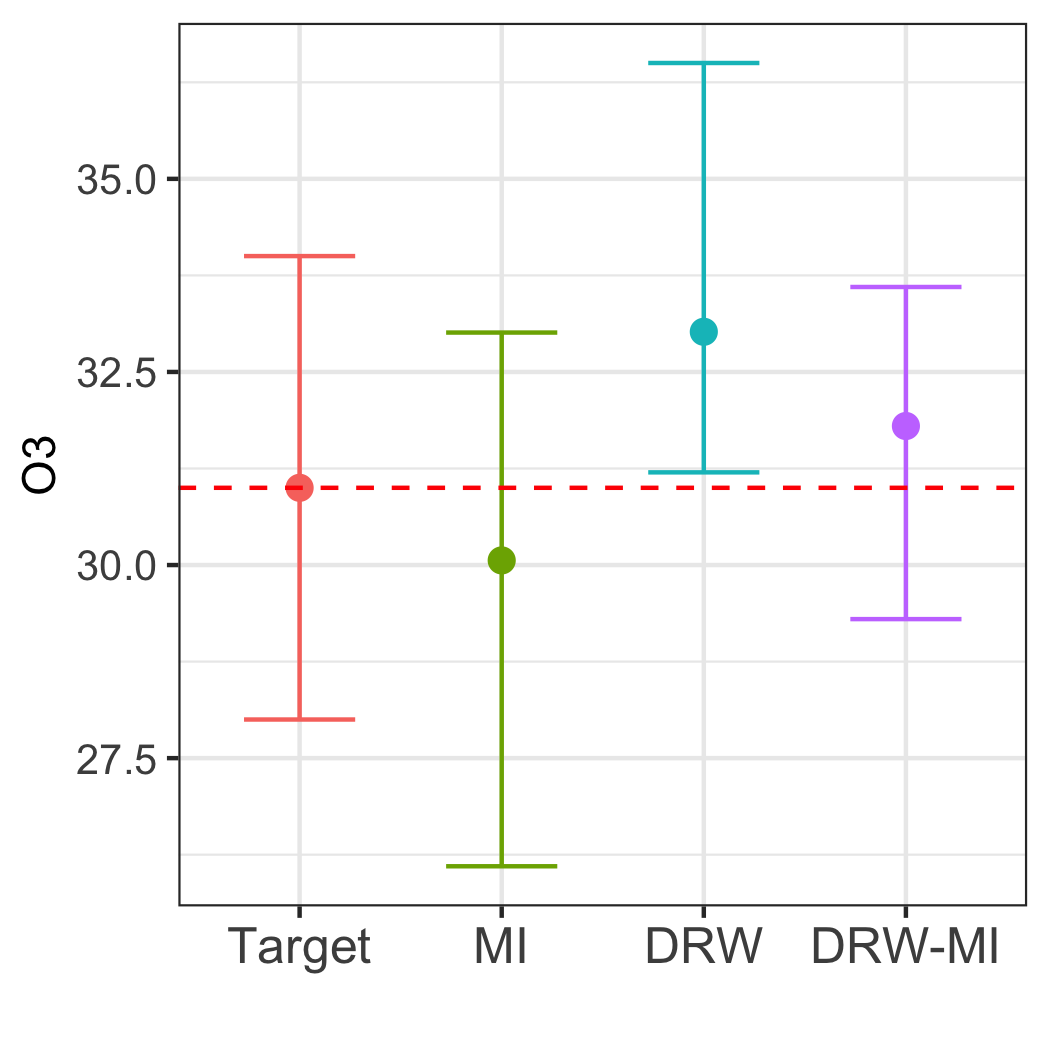}
\end{subfigure}
\begin{subfigure}[t]{0.24\textwidth}
    \centering
    \caption{$50\%$-quantile}
    \includegraphics[width=0.9\textwidth]{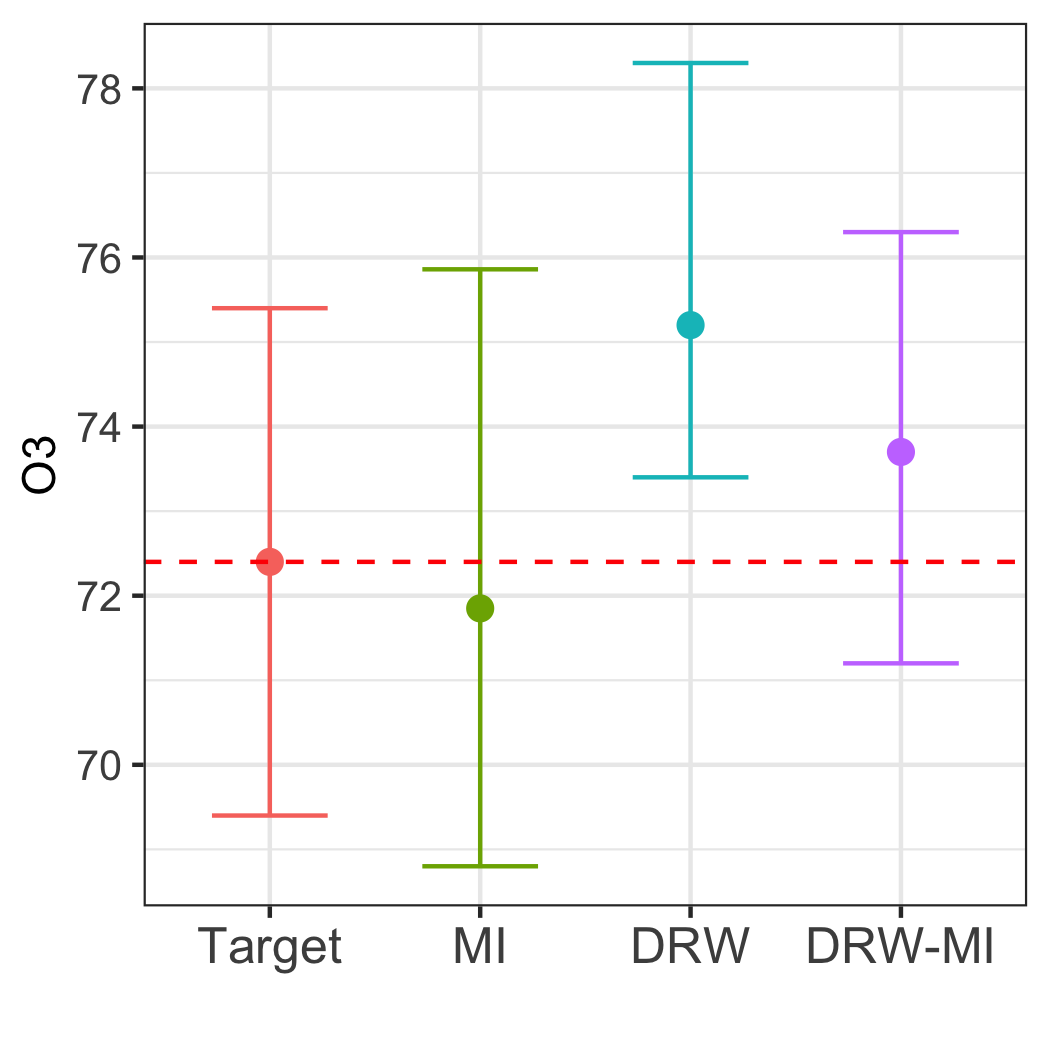}
\end{subfigure}
\begin{subfigure}[t]{0.24\textwidth}
    \centering
    \caption{$75\%$-quantile}
    \includegraphics[width=0.9\textwidth]{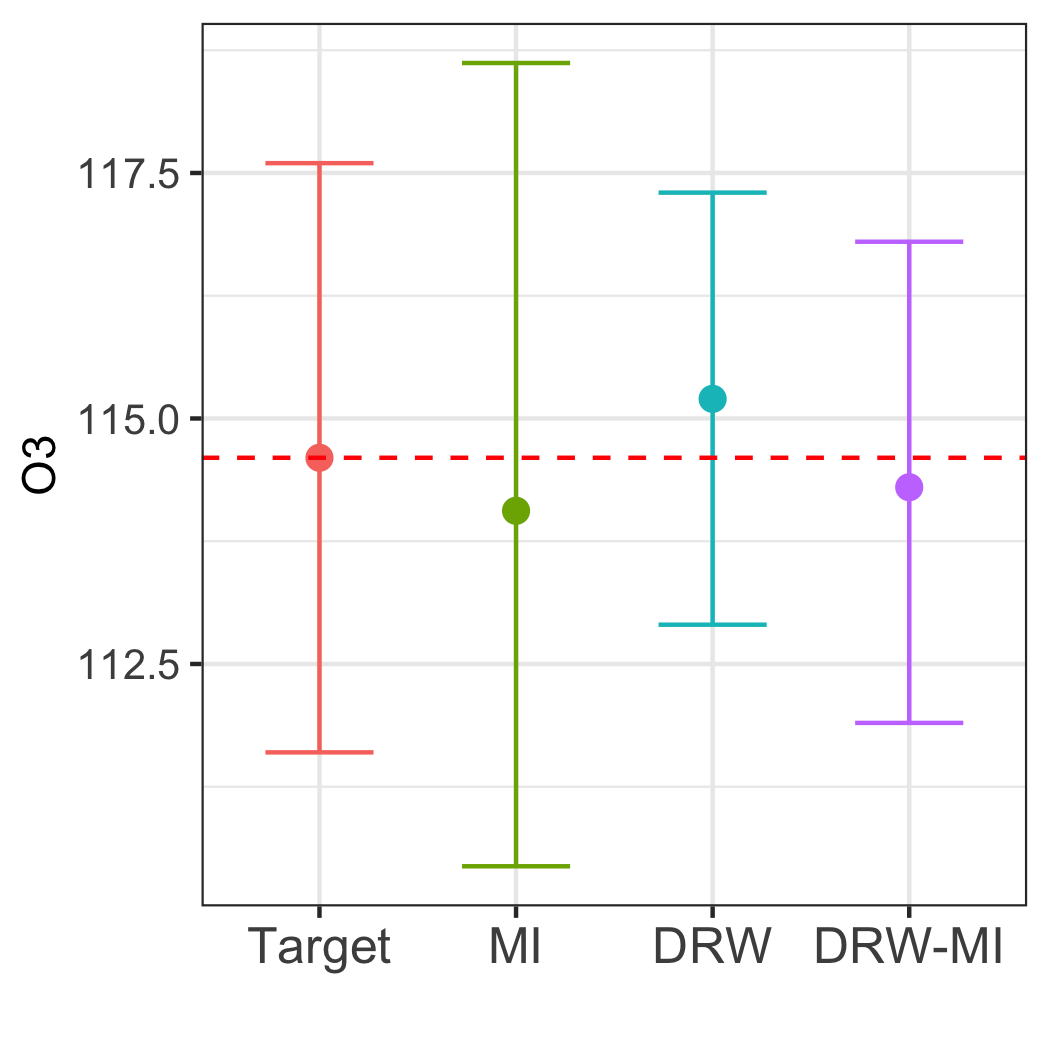}
\end{subfigure}
\label{fig: O3-inference}
\end{figure}

\section{Discussion}
\label{sec: discuss}

This study investigates the statistical inference for general estimating equations with the covariate shift transfer learning. Instead of the common strategy of density ratio weighting, we construct an orthogonal estimating equation that is more robust against nuisance function estimation errors. To address the challenge that the conditional mean estimating function is parameter-dependent, we adopt a multiple-imputation approach that avoids conducting the regression at infinitely many parameters. Our estimation for the nuisance functions accommodates flexible uses of ML algorithms. The theoretical results reveal that the EL estimator based on the orthogonal estimating equation is semiparametric efficient. 
Compared with the related literature such as \cite{chen2024}, the inference does not require a Bootstrap procedure, as it is shown that the log El ratio restores the Wilks theorem, despite the presence of nuisance functions. We also discuss the DNN-based nuisance function estimation to alleviate the curse of dimensionality. 

There are some intersting extensions that may be considered in subsequent research. First, in this work we focus on the transfer learning under the covariate shift
Investigation on
how to handle general distribution shift settings, such as the label shift and domain generalization, remains an important
avenue for future work. Second, the density ratio function is required to be uniformly bounded, as a common assumption for nonparametric estimation. However, such a condition can be violated in scenarios where the target and the source domain have non-overlap regions, or the proportion of the target sample in the full sample converges to $0$, namely $n / N \to 0$ as $N \to \infty$. Such scenarios have been considered in studies of semi-supervised learning, such as \cite{zhang2022high}  and \cite{chakrabortty2022general}. However, the current TL setting is more challenging due to the covariate shift, where the density ratio function should be estimated. 
In addition, while we have investigated the scenario where the dimension grows with the sample size in Section \ref{sec: hd}, it does not accommodate the ultra-high dimension regimes where the dimension can be larger than the sample size. Extensions to the ultra-high dimensional GEEs under transfer learning can possibly be achieved with the penalized EL established by \cite{changNewScopePenalized2018}, while the effect of nuisance function estimation should be carefully examined.

\bibliographystyle{apalike}

\bibliography{Main_ref}

\end{document}